\documentclass[10pt,english,aps,amssymb,prb,twocolumn, showpacs]{revtex4-1}
\pdfoutput=1
\usepackage[utf8]{inputenc}
\usepackage{amsmath}
\usepackage{graphicx}
\usepackage{amssymb}
\usepackage{esint}
\usepackage{verbatim}

\makeatletter
\@ifundefined{textcolor}{}
{
 \definecolor{WHITE}{gray}{1}
 \definecolor{RED}{rgb}{1,0,0}
 \definecolor{GREEN}{rgb}{0,1,0}
 \definecolor{BLUE}{rgb}{0,0,1}
 \definecolor{CYAN}{cmyk}{1,0,0,0}
 \definecolor{MAGENTA}{cmyk}{0,1,0,0}
 \definecolor{YELLOW}{cmyk}{0,0,1,0}
 }

\newcommand{\bra}[1]{\langle #1|}
\newcommand{\ket}[1]{|#1\rangle}

\newcommand{\brakets}[2]{\left\langle#1| #2 \right\rangle}
\newcommand{\IM}{\text{Im}}
\newcommand{\RE}{\text{Re}}
\renewcommand{\phi}{\varphi}
\renewcommand{\epsilon}{\varepsilon}
\newcommand{\eff}{\textrm{eff}}
\renewcommand{\vec}[1]{{\bf #1}}

\newcommand{\tr}{\text{tr}}
\newcommand{\sgn}{\text{sgn}}
\newcommand{\bs}{\boldsymbol}
\newcommand{\mc}{\mathcal}

\usepackage{babel}

\begin{document}
\title {Topological superconductivity in ferromagnetic atom chains beyond the deep-impurity regime}
\author{Kim Pöyhönen}
\author{Alex Westström}
\author{Teemu Ojanen}
\email[Correspondence to ]{teemuo@boojum.hut.fi}
\affiliation{Department of Applied Physics (LTL), Aalto University, P.~O.~Box 15100,
FI-00076 AALTO, Finland }
\date{\today}
\begin{abstract}
Recent developments in the search for topological superconductivity have brought lattices of magnetic adatoms on a superconductor into intense focus. In this work we will study ferromagnetic chains of  adatoms on superconducting surfaces with Rashba spin-orbit coupling. Generalising the deep-impurity approach employed extensively in previous works to arbitrary subgap energies, we formulate the theory of the subgap spectrum as a nonlinear matrix eigenvalue problem. We obtain an essentially analytical description of the subgap spectrum, allowing an efficient study of the topological properties. Employing a flat-band Hamiltonian sharing the topological properties of the chain, we evaluate the $\mathbb{Z}$-valued winding number and discover five distinct topological phases. Our results also confirm that the topological band formation does not require the decoupled Shiba energies to be fine-tuned to the gap centre. We also study the properties of Majorana bound states in the system.

\end{abstract}
\maketitle
\bigskip{}

\section{introduction}

The physics of magnetic impurities in superconductors has been studied extensively since the seminal works by Yu, Shiba and Rusinov.\cite{yu:1965:1,shiba:1968:1,rusinov:1969:1} A magnetic moment locally disrupts the superconducting condensate, giving rise to subgap bound states localised in the vicinity of the magnetic moment. In the past two decades, Scanning Tunneling Microscopy (STM) studies of magnetic impurity physics have complemented this theoretical understanding.\cite{balatsky:2006:1, yazdani:1997:1} The interest in these systems was renewed after proposals\cite{choy:2011:1,nadj-perge:2013:1,pientka:2013:1,vazifeh:2013:1,klinovaja:2013:1,heimes:2014:1,brydon:2015:1} to realise 1D and 2D topological superconductors by lattices of magnetic adatoms on superconducting surfaces. The promising observation\cite{nadj-perge:2014:1} of signatures of Majorana bound states (MBSs) in ferromagnetic chains has brought the topic in the intense focus recently. Similar signatures are also claimed to have been observed in another recent experiment.\cite{pawlak:2015:1} Magnetic chains, with their respective advantages and weaknesses, offer an interesting alternative to the nanowire systems\cite{oreg:2010:1,lutchyn:2010:1,mourik:2012:1,das:2012:1} as a route towards topological superconductivity.   

In this work we concentrate on the properties of 1D ferromagnetic adatom lattices deposited on a 2D superconducting film with a Rashba spin-orbit coupling (SOC). We will study the situation where each magnetic moment binds a single subgap Shiba state with energy $\Delta\frac{1-\alpha^2}{1+\alpha^2}$, where the dimensionless coupling $\alpha=JS\nu_0\pi$ is determined by microscopic parameters -- $\nu_0$ is the density of states, $J$ is the exchange coupling and $S$ is the magnitude of the impurity spin -- and $\Delta$ is the pairing gap of the superconductor.\cite{pientka:2013:1,brydon:2015:1} The Shiba states are concentrated around the magnetic moments with wavefunctions decaying as $\frac{e^{-r/\xi}}{r^{1/2 } }$ in 2D and $\frac{e^{-r/\xi}}{r }$ in 3D superconductors, where the exponential decay is controlled by the superconducting coherence length $\xi_0$. The wavefunctions are more slowly decaying in 2D, which was qualitatively confirmed in a recent experiment.\cite{menard:2015:1} The slow spatial decay of the bound states at length scales below $\xi$ has important physical implications. The separation of the magnetic moments can easily be much shorter than the coherence length, thus allowing a significant hybridisation of bound states separated by  dozens of neighbours. Therefore effective theories of subgap bands generically exhibit a long-range hopping, adding its special properties to the problem. The long-range hopping nature of the Shiba states will give rise to rich topological properties, especially in 2D magnetic lattices where it is possible to achieve dozens of different phases and Chern numbers much larger than unity.\cite{rontynen:2015:1,li:2015:1} 

\begin{figure}
\includegraphics[width=0.95\linewidth]{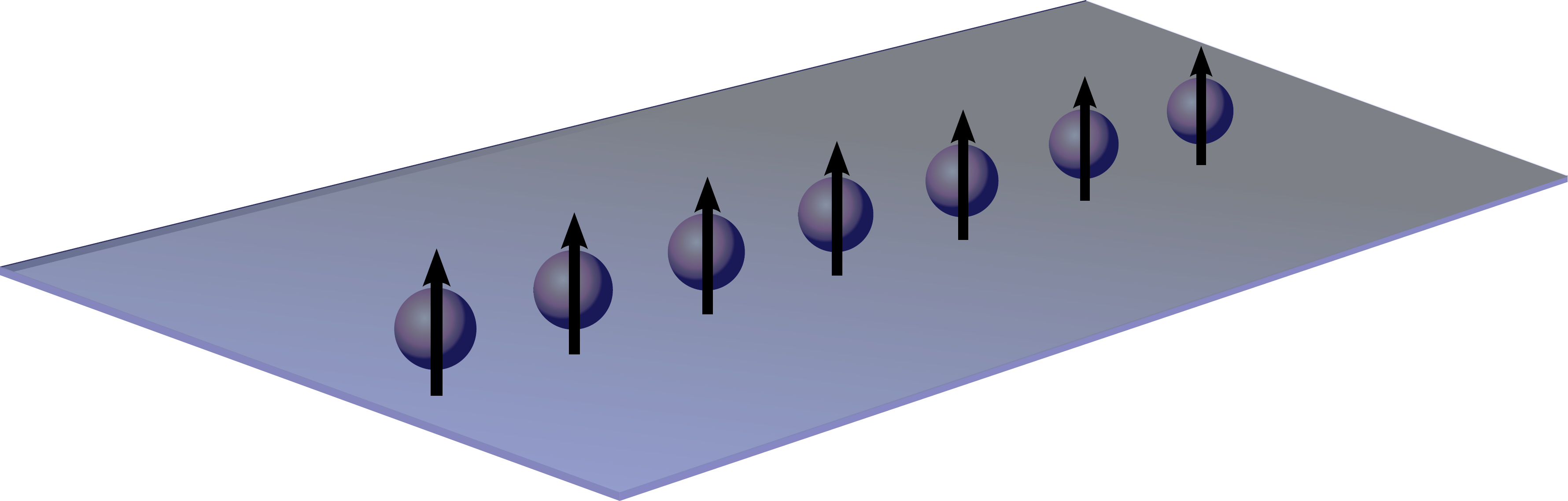}
\caption{Schematic figure of the ferromagnetic chain of atoms embedded on a two-dimensional superconductor.}\label{fig:schematic}
\end{figure}

The topological properties of magnetic chains with helical\cite{nadj-perge:2013:1,pientka:2013:1,pientka:2014:1,poyhonen:2014:1,rontynen:2014:1,reis:2014:1,vazifeh:2013:1,braunecker:2013:1,klinovaja:2013:1} and ferromagnetic\cite{heimes:2015:1, li:2014:2, zhang:2015:1,bjornson2015:1} order have already been studied in considerable detail. An interesting recent step towards a more realistic description of Shiba systems was taken by the generalisation of the theory of topological Shiba bands to the multichannel case to accommodate multiple bound states on a single impurity.\cite{zhang:2015:1} That work, like most of the studies capturing the microscopic structure of the Shiba states, is restricted to the deep-dilute impurity limit where the energy of the decoupled Shiba states reside close to the gap centre, i.e. $\alpha\approx 1$. This is motivated partly by the significant theoretical simplification arising from the linearisation of the spectral problem in energy, which allows an elegant and simple formulation in terms of an effective Hamiltonian.\cite{pientka:2013:1}  Our work takes a step towards a more general description by formulating a theory of Shiba bands beyond the deep-impurity regime. This is important, since tuning the parameters controlling the value of $\alpha$ is difficult in experimental setups. As highlighted in our work, the microscopic parameters do not need to be fine tuned to the deep-impurity regime to reach the topological phase. 

In Sec.~\ref{model}, we formulate the subgap spectral problem of a ferromagnetic Shiba chain as a nonlinear matrix eigenvalue problem and obtain an essentially analytical solution for the dispersion and wavefunctions valid for arbitrary subgap energies. In Sec.~\ref{topo}, we show that the 1D chain has an emergent chiral symmetry which enables us to defining a $\mathbb{Z}$-valued winding number invariant $\mc N$ classifying the different phases. In Sec.~\ref{sec:phasediagrams}, we discover five distinct topological phases with winding numbers $\mc N=-2,-1,0,1,2$, discuss the implications of the observation and compare the results to the deep-impurity formulation. We study the MBSs supported by the chain in Sec.~\ref{MBS}, extracting the spatial decay of their wavefunctions, topological gaps and energy splitting. In Sec.~\ref{summ}, we summarise our results and discuss their implications. 

\section{Model and the spectral problem}\label{model}

\subsection{ Subgap spectrum as a nonlinear eigenvalue problem}
The starting point of our study is the system depicted in Fig.~\ref{fig:schematic} consisting of a 2D superconductor surface decorated by magnetic impurities arranged in a chain with lattice constant $a$. The system is described by the Bogoliubov-de Gennes Hamiltonian
\begin{equation}\label{eq:hamiltonian}
\begin{split}
\mathcal{H} = \left(\frac{k^2}{2m} - \mu + \alpha_R(k_y\sigma_x - k_x\sigma_y)\right)\tau_z \\+ \Delta\tau_x - J\sum_i(\vec{S}\cdot\bs\sigma_i)\delta(\vec{r} - \vec{r}_i).
\end{split}
\end{equation}
The matrix structure of Eq.~(\ref{eq:hamiltonian}) corresponds to the Nambu basis  $\Psi = (\psi_{\uparrow}, \psi_{\downarrow}, \psi^\dag_{-\downarrow}, -\psi^\dag_{\uparrow})^T$; $\tau_i$ and $\sigma_i$ correspond to Pauli spin matrices in the particle-hole and spin subspaces, respectively. Here $k^2/2m - \mu$ is the kinetic energy of the electrons, $\alpha_R$ is the spin-orbit coupling and $\Delta$ describes the pairing amplitude of Cooper pairs. The vector $\vec r$ is the position of the electron, whereas $\vec r_i$ describes the impurity positions. We will focus on the case where all magnetic moments point in the $z$ direction. Consequently the system is in the symmetry class D.\cite{schnyder:2009:1,ryu:2010:1} Inserting the Hamiltonian density into the Bogoliubov-de Gennes equation $\mc H \Psi = E\Psi$, we obtain
\begin{equation}\label{eq:postBdG}
\begin{split}
\left[E-\left(\xi_{\vec k} + \alpha_R(k_y\sigma_x - k_x\sigma_y)\right)\tau_z - \Delta\tau_x\right]\Psi(\vec r) \\= - J\sum_i S\sigma_z\delta(\vec{r} - \vec{r}_i)\Psi(\vec r_i)
\end{split}
\end{equation}
where we have introduced $\xi_{\vec k} = k^2/(2m) - \mu$. We can make further progress by restricting the system to a one-dimensional chain in the $x$ direction, yielding, as seen in Appendix A, the equation
\begin{equation}\label{eq:rashbaeq}
\begin{split}
\left(1 - \alpha\frac{E + \tau_x}{\sqrt{\Delta^2 - E^2}}S\sigma_z\right)\Psi(x_i) \\= -\sum_{i\neq j} J_E(x_{ij})S\sigma_z\Psi(x_j).
\end{split}
\end{equation}
where $\alpha = JS\nu_0\pi = \frac{1}{2}JSm$, and $J_E$ is 
essentially given by the propagator of the 2D bulk electrons. The specific form of $J_E$ in this case was first evaluated by Brydon \textit{et al.} \cite{brydon:2015:1} and is presented in Appendix A. In the deep-dilute limit this equation can be linearised in $E$ and $1/\sqrt{k_Fa}$, where $k_F$ is the Fermi wave number and $a$ is the chain lattice constant; this allows projection onto the low-energy single impurity bands, resulting in an effective two-band Hamiltonian.\cite{pientka:2013:1} However, in this work we are interested in the behaviour of the system for arbitrary subgap energies, thus requiring a more general approach to the problem. As outlined in Appendix A, it is convenient to work in the basis of the eigenstates of $\tau_x\sigma_z$, in which case we can obtain a nonlinear eigenvalue problem (NLEVP) for $\lambda\equiv (\Delta+E)/\sqrt{\Delta^2-E^2}$ and $\Psi$,
\begin{equation}\label{eq:nlevp}
\begin{pmatrix}
a\lambda^2-\tfrac{1}{\alpha}\lambda & b\lambda & c\lambda^2 & -\lambda d\\
-b\lambda & \tfrac{1}{\alpha}\lambda - a & -\lambda d & c\\
-c\lambda^2 & -\lambda d & a\lambda^2 + \tfrac{1}{\alpha}\lambda & -b\lambda\\
-\lambda d & -c & b \lambda & -\tfrac{1}{\alpha}\lambda - a
\end{pmatrix}
\Psi = 0.
\end{equation}
Here $a, b, c, d$ are $N\times N$ matrices (where $N$ is the length of the chain), and similarly $\lambda$ is shorthand for $\lambda \mathbb{I}_{N\times N}$. The coefficient matrices are of the form
\begin{equation}
\begin{split}
a_{ij} &= -\frac{\sqrt{\Delta^2-E^2}}{2m}(I_3^-(x_{ij}) + I_3^+(x_{ij})) + \delta_{ij}\\
b_{ij} &= - \frac{i}{2m}(I_2^-(x_{ij}) - I_2^+(x_{ij}))\\
c_{ij} &= -i\frac{\sqrt{\Delta^2-E^2}}{2m}(I_4^-(x_{ij}) - I_4^+(x_{ij}))\\
d_{ij} &= - \frac{1}{2m}(I_1^-(x_{ij}) + I_1^+(x_{ij})),
\end{split}
\end{equation}
where $x_{ij} \equiv x_i - x_j$ denotes the difference between positions of magnetic moments and $I^{\pm}_i$ are given by the special function expressions presented in Appendix A. It is important to note that the submatrices have a nontrivial energy dependence through the energy dependent coherence length $\xi_E \equiv v_F/\sqrt{\Delta^2-E^2}$ (not to be confused with the kinetic energy $\xi_\mathbf{k}$), which in principle complicates the solution considerably. In the limit where the energy-independent coherence length $\xi_0 \equiv v_F/\Delta$ goes to infinity this nontrivial dependence vanishes. The problem  (\ref{eq:nlevp}) is a representative of NLEVPs $A(E)\Psi=0$, where $A(E)$ is a matrix-valued non-linear function of $E$ and can be thought of as a generalisation of the usual linear Schrödinger problem where $A(E)=H-E\mathbb{I}$. 

On physical grounds we know that since we are dealing with a system of $N$ magnetic moments with a single $s$-channel bound state, Eq.~(\ref{eq:nlevp}) has  $N$ positive and $N$ negative energy solutions describing the magnetic subgap band.  Most of the previous works solve the problem in the deep-impurity limit by linearising the model, yielding a linear problem $H^\eff\Psi = E\Psi$  with an effective  $2N\times 2N$ Hamiltonian
\begin{equation}\label{eq:h2band}
H^{\eff} =
\Delta\begin{pmatrix}
\mathbb{I} -\alpha\tilde{a} & i\alpha\tilde{c}\\
-i\alpha\tilde{c}^\dagger & \alpha\tilde{a}-\mathbb{I}
\end{pmatrix}.
\end{equation}
where $\tilde{a} = \lim_{E\to0}a$, $\tilde{c} = \lim_{E\to0}c$. This description is valid for deep impurities $\alpha\approx1$ close to the gap centre coupled by a weak hopping $1/\sqrt{k_Fa}\ll1$. One of our central results presented below is that by relaxing the conditions leading to the two-band model we can show that the topologically nontrivial phases extend outside the deep-impurity regime. Therefore it is not necessary for the decoupled Shiba states to lie close to the gap centre to achieve topological phases, and there is no need to tune their energies if the hybridisation, which can be controlled by the distance between the magnetic moments, is sufficiently strong.

\subsection{Solution to the nonlinear eigenvalue problem}
In the case of a periodic or infinite system, the problem can be vastly simplified by going over to reciprocal space. The submatrices $a,b,c,d$ are all translationally invariant, allowing a Fourier transform
\begin{equation}\label{eq:fourier}
a_k = \sum_{j\neq 0} a_{ij}e^{ika(i-j)}
\end{equation}
which reduces the equation to a $4\times 4$ NLEVP. In this form we can find an expression for the bulk spectrum from the usual eigenvalue condition by requiring that the determinant of the matrix in Eq.~(\ref{eq:nlevp}) vanishes. This gives us a transcendental equation for the energy,
\begin{widetext}
\begin{equation}\label{eq:energy}
E(k)= \pm\Delta\sqrt{\frac{(a_k^2 + b_k^2+c_k^2+d_k^2-1/\alpha)^2-4(a_kb_k+c_kd_k)^2}{(a_k^2 + b_k^2+c_k^2+d_k^2-1/\alpha)^2-4(a_kb_k+c_kd_k)^2+4(a_k^2+c_k^2)}}.
\end{equation}
\end{widetext}
In comparison, the expression for the energy of the linearised two-band model is
\begin{equation}
E^\eff(k) = \pm\Delta\sqrt{(1-\alpha \tilde a_k)^2 + \alpha^2\tilde c_k^2}
\end{equation}
-- note that this is not a transcendental equation, because $E$ is set to 0 in the expressions for $a_k$ and $c_k$ and hence it gives $E$ directly for any coherence length. However, Eq.~(\ref{eq:energy}) can be solved numerically for any coherence length with standard methods for transcendental equations. As shown in Appendix B, it turns out that replacing $\xi_E$ with $\xi_0$ in the four-band model gives an excellent approximation of the energy. This has two important consequences. Firstly, after setting $\xi_E\to\xi_0$ on the right-hand side, Eq.~(\ref{eq:energy}) essentially provides an explicit solution of the subgap energy bands typically within accuracy of the order of $10^{-2}\Delta$ or even much better. The accuracy of the solution could be numerically improved systematically by iteration as also shown in Appendix B, but for all practical purposes the zeroth order solution is sufficient. Secondly, since the replacement $\xi_E\to \xi_0$ in the bulk spectral problem is seen to lead to negligible corrections to the energies, we also employ this approximation when we solve the NLEVP in real space to study the MBSs. 

Remarkably, the eigenspinor can also be solved analytically from the $4\times 4$ NLEVP in reciprocal space. The components of the eigenspinor $\Psi_\lambda(k) \equiv (x_1,\,x_2,\,x_3,\,x_4)^T$ are, up to a normalisation factor,
\begin{equation}\label{eq:components}
\begin{split}
x_1 &= -\lambda\left[c^2-d^2 + \frac{\lambda^2d^2-(ad-bc)^2}{\lambda^2-(a^2+c^2)} \right]\\
x_3 &= \lambda(ac-bd) - c + \frac{\lambda(ab+cd+\lambda b)(bc-ad+\lambda d)}{\lambda^2-(a^2+c^2)} \\
x_2 &= \frac{\lambda(ab+cd+\lambda b)}{\lambda^2-(a^2+c^2)}x_1 + \frac{\lambda(ad-bc+\lambda d)}{\lambda^2-(a^2+c^2)}x_3\\
x_4 &= \frac{1}{c}\left[\lambda b x_1 + (a-\lambda)x_2 + \lambda d x_3\right].
\end{split}
\end{equation}
where we have suppressed the $k$ index on the functions $a,b,c,d$. The eigenvector $|E_+\rangle$ corresponding to the positive energy can be found by inserting the appropriate solution $E(k)>0$  of Eq.~(\ref{eq:energy}) into the expressions (\ref{eq:components}) for the components. Importantly, the negative energy solution $|E_-\rangle$ corresponding to $E(k)<0$ is related to the positive energy solution (at the same $k$) through $\mc C|E_+\rangle=|E_-\rangle$  where $\mc C = \tau_y\sigma_y$ in the original basis of Eq.~(\ref{eq:hamiltonian}), or $\sigma_z\otimes\sigma_x$ in the rotated basis in which solution (\ref{eq:components}) is obtained.  The expression $\sigma_z\otimes\sigma_x$ stands for a $4\times 4$ matrix where the Pauli matrices do not refer to the electron spin anymore.

\section{Topological considerations}\label{topo}

\subsection{1D flat-band Hamiltonian with chiral symmetry}
In this section we will present the topological properties of the subgap Shiba bands described by Eqs.~(\ref{eq:energy}) and (\ref{eq:components}). In the periodic table of topological insulators and superconductors,  $d$-dimensional gapped systems are typically classified by studying associated $d$-dimensional (Bloch) Hamiltonians and their symmetries. However, while we started with a 2D Hamiltonian (\ref{eq:hamiltonian}) we obtained subgap energy bands that are essentially localised in the vicinity of the 1D chain of magnetic atoms. In addition, the eigenstates $|E_{\pm}\rangle$ are parametrised by the 1D wave vector $k$ and so the subgap spectrum is effectively one-dimensional. In the deep-dilute limit it is possible to reduce the NLEVP to a linear problem in terms of an effective Hamiltonian of type (\ref{eq:h2band}) and study the topological properties by treating $H^{\eff}$  as a \emph{bona fide} 1D Hamiltonian. However, in our case another strategy must be adopted due to the nonlinear formulation of the problem. Therefore we define an effective Hamiltonian in terms of the projectors of the subgap bands as   
\begin{equation}\label{eq:htilde}
\tilde{H} = \sum_{\nu=\pm} E_\nu\ket{E_\nu}\bra{E_\nu} \equiv \sum_{\nu=\pm}E_\nu \hat{P}_\nu = E_+ \hat{P}_+ + E_- \hat{P}_-.
\end{equation}
The low-energy properties of the initial 2D system coincide with those of the 1D Hamiltonian (\ref{eq:htilde}). When we are only interested in the topological properties of the system, the precise details of the energy bands are irrelevant and with no loss of generality we can adiabatically flatten them $E_\pm(k)\to\pm1$ and study $\tilde{H} = \hat{P}_+ - \hat{P}_-$.  Notice that the property $\mc C|E_+\rangle=|E_-\rangle$ implies the anticommutation relation $\{\mc C,\tilde{H} \}=0$, i.e., the 1D system has the chiral symmetry and belongs to the symmetry class BDI\cite{schnyder:2009:1,ryu:2010:1} as the original Hamiltonian (\ref{eq:hamiltonian}) if motion is restricted to the $x$ direction ($k_y=0$).

\subsection{Winding number}
In noninteracting 1D systems with chiral symmetry the topological classification is $\mathbb Z$-valued, thus the system can support an arbitrary number of different topological states.  
These states can be most conveniently obtained by evaluating the winding number invariant given in the form
\begin{equation}\label{eq:winding}
\mathcal N = \frac{1}{4\pi i}\int_{-\pi/a}^{\pi/a} \mathrm{d}k\ \tr\left[\mathcal C H^{-1}\partial_k H\right].
\end{equation}
where $\mathcal C = \mathcal C^\dagger = \mathcal C^{-1}$ is the chiral symmetry operator and $H$ is a fully gapped 1D Bloch Hamiltonian.\cite{schnyder:2009:1,ryu:2010:1} As discussed above, in the transformed basis the chiral symmetry operator for this system is $\mc C = \sigma_z\otimes\sigma_x$.  After inserting the flat-band Hamiltonian $\tilde{H}$ in Eq.~(\ref{eq:winding}) and a few lines of algebra, outlined in Appendix C, the winding number can be expressed in terms of the band projectors as 
\begin{equation}\label{eq:finalwinding}
\mathcal N= \frac{1}{\pi i}\int_{-\pi/a}^{\pi/a} \mathrm{d}k\ \tr\left[\mathcal C \hat{P}_+\partial_k \hat{P}_+ \right].
\end{equation}
In the next section, we will analyse the topological phase diagram by evaluating (\ref{eq:finalwinding}) as a function of the system parameters. 
 
\subsection{$\mathbb{Z}_2$ invariant}
The existence of chiral symmetry is typically approximate in real systems and topological classification based on particle-hole redundancy that is built into the Bogoliubov-de Gennes formalism is expected to be less sensitive to disorder. Therefore we also consider Kitaev's $\mathbb{Z}_2$ invariant measuring the ground state fermion parity. 
 The phase transition between different parities is associated by the energy gap closing $E(k)=0$ at $k=0,\pi/a$. This is useful, as the phase boundary between of different $\mathbb{Z}_2$ phases can be explicitly obtained from the NLEVP matrix in reciprocal space, as discussed in Ref.~\onlinecite{weststrom:2015:1}.  We are especially interested in the phase diagram as a function of the single-impurity Shiba energy determined by $\alpha$ and the hybridisation parameter $k_Fa$. Setting $E(k)=0$ at $k=0,\pi/a$, we can solve the gap closing lines  as a function of $\alpha$ and $k_Fa$ as
\begin{equation}\label{eq:teemutrick}
\alpha = \frac{1}{\sqrt{\tilde a_k^2+\tilde d_k^2}}\bigg|_{k=0,\pi/a}.
\end{equation}
This equation defines the phase boundary between regions of different parity of the $\mathbb{Z}_2$ invariant, and is valid for arbitrary coherence lengths. The corresponding equation for the two-band model is given by
\begin{equation}\label{eq:teemutrick2b}
\alpha = \frac{1}{\tilde a_k}\bigg|_{k=0,\pi/a},
\end{equation}
allowing for simple comparison of the $\mathbb{Z}_2$ phase diagram of the exact four-band solution and the two-band approximation valid in the deep-impurity limit. In the presence of chiral symmetry the $\mathbb{Z}_2$ phases are captured by the winding number parity so considering it separately brings little new information. However, formulas (\ref{eq:teemutrick}) and (\ref{eq:teemutrick2b}) yield semianalytic phase boundaries of the different parity phases and make their analysis very convenient without the need to perform integral (\ref{eq:finalwinding}) everywhere. 

\begin{figure}
\includegraphics[width=0.95\linewidth]{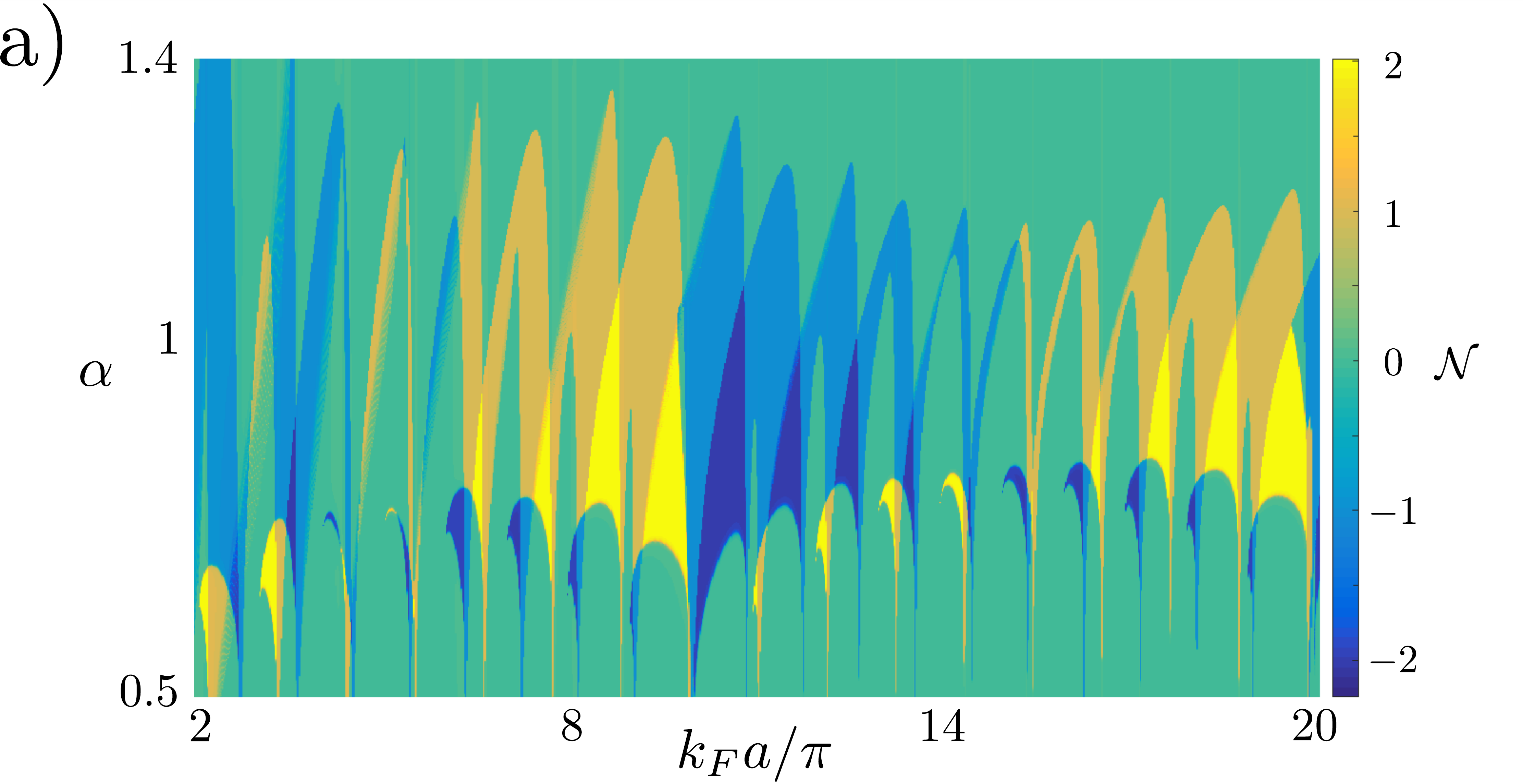}
\includegraphics[width=0.95\linewidth]{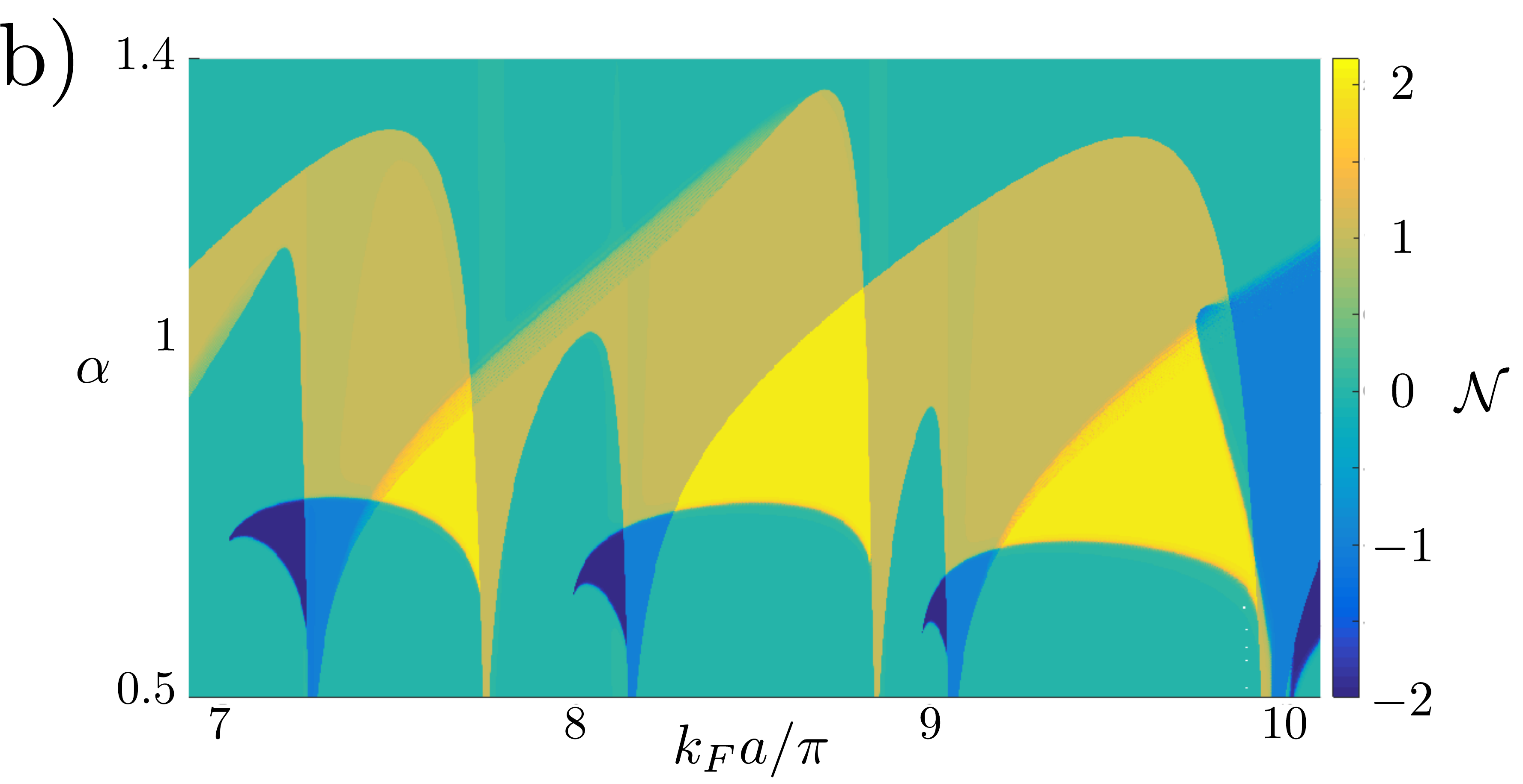}
\includegraphics[width=0.95\linewidth]{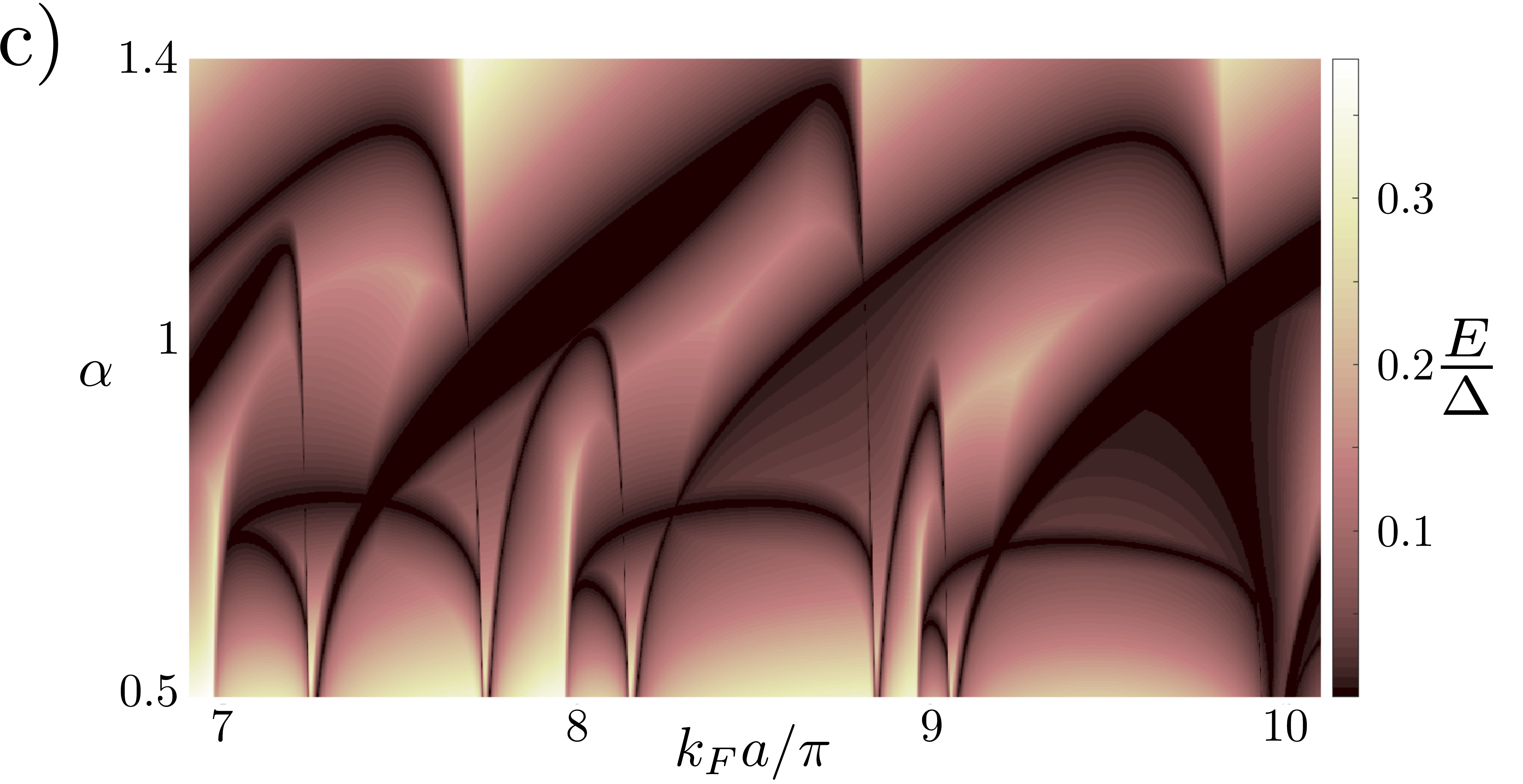}
\caption{a) Topological diagram of the four-band model, created by plotting the winding number as a function of $k_Fa$ and $\alpha$. The parameters used are $\xi_0 = 50a$, $\varsigma = 0.1$. Winding numbers from $-2$ to $+2$ are represented. As $k_Fa$ increases the winding number configurations (quasi-)periodically flip signs; this happens more often with increasing $\varsigma$. b) Same parameters as in a), but focused on a narrower range of $k_Fa$ values. c) Energy of the lowest-lying positive state for an infinite system as a function of $k_Fa$ and $\alpha$. Parameters used are the same as in a). The topological phase transitions are seen here as bulk gap closings.  }\label{fig:winding}
\end{figure}

\section{Topological phase diagrams}\label{sec:phasediagrams}

In the previous section we derived analytical formulas describing the FM Shiba chain. In this section we will use those results to examine the topological properties of the chain. To that end we have used Eqs.~(\ref{eq:finalwinding}) and (\ref{eq:teemutrick}) to calculate the topological phase diagram of the system.

As seen in Fig.~\ref{fig:winding}, the phase diagram thus obtained reveals the presence of several topological phases, in accordance with the BDI classification of the FM chain. In total we found five phases, corresponding to winding numbers $0, \pm 1, \pm 2$. This feature is present in both the general model and in the linearised two-band model, a feature missed in previous works which either considered the parity of the winding number\cite{brydon:2015:1} or examined a parameter area too narrow to contain all the winding numbers.\cite{heimes:2015:1} Often, nontrivial regions appear for parameter values where the validity of the two-band model breaks down, so a reliable identification of the phase diagram requires the full machinery of the four-band model derived in the present work. As seen in Fig.~\ref{fig:winding} c), the topological phase transitions are accompanied by closings of the gap as expected. A phase with winding number $\mathcal{N}$ supports $|\mathcal{N}|$ degenerate MBSs; further, domain walls between regions with different winding numbers $\mathcal{N}_1$ and $\mathcal{N}_2$ can support $|\mc N_1 - \mc N_2|$ bound states. However, while the energy gaps for phases  $|\mathcal{N}|=1$ can be as high as $0.25\Delta$, the  $|\mathcal{N}|=2$ phases generally have smaller gaps; this naturally provides a complication in realising domain walls with more than two MBSs. The general behaviour of the energy gap will be discussed in more detail in the next section.
Notice that significant regions of the nontrivial topological phases extend to values $\alpha>1.5$ and $\alpha<0.6$ for small values of  $k_Fa$ (large hybridisation), translating to bound state energies between $-0.4\Delta$ and $0.5\Delta$. Thus, provided that the separation of the moments can be tuned in the fabrication of the chain, \emph{there is no need for tuning the bound state energies close to the gap centre}.  

\begin{figure}
\includegraphics[width=0.9\linewidth]{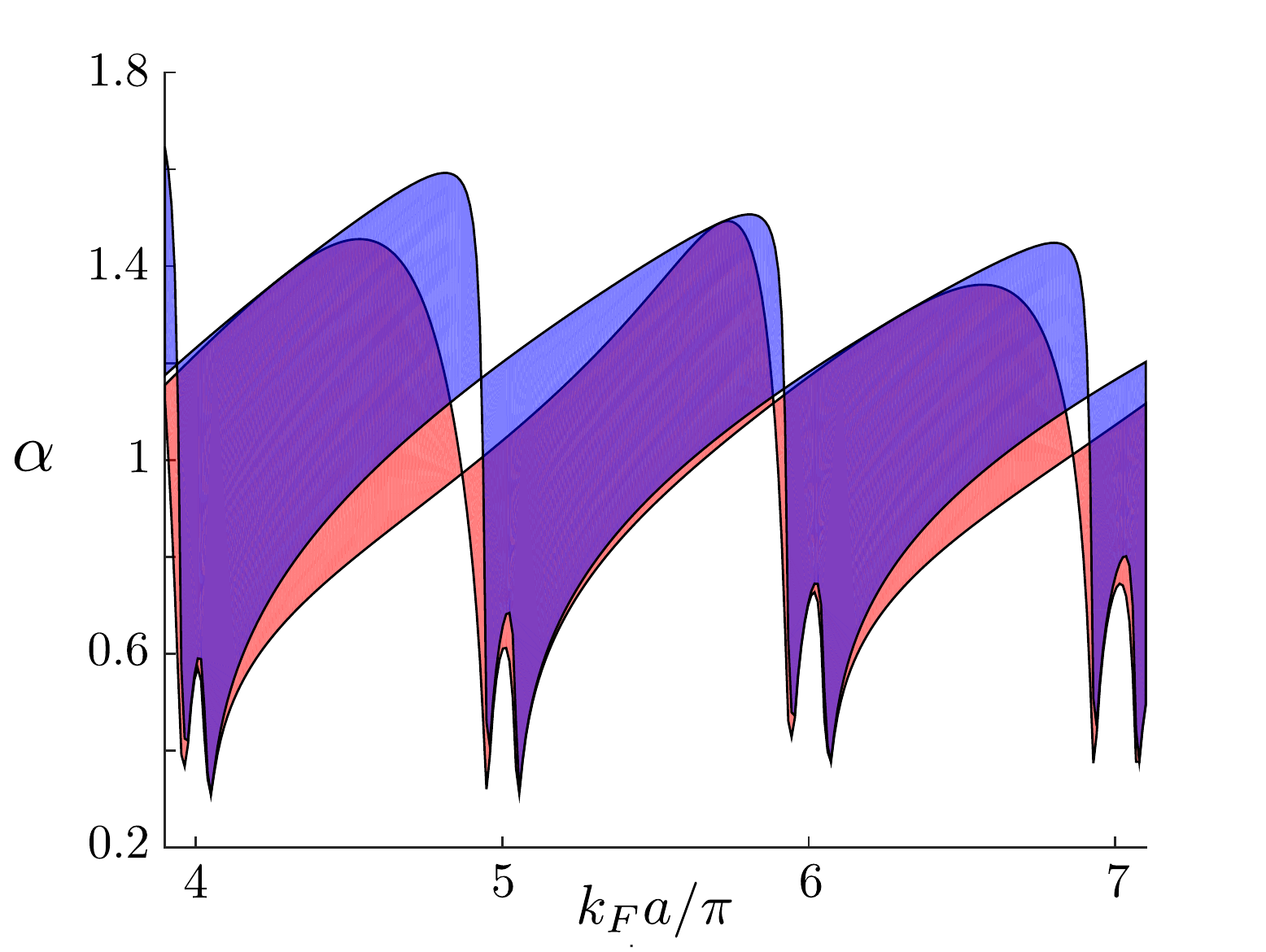}
\caption{$\mathbb{Z}_2$ phase diagram of the exact four-band model (red) and the two-band deep-impurity (blue) model. Parameters used are $\xi_0 = 50a$, $\varsigma = 0.01$. Note that the agreement is not accurate even in the deep-impurity $\alpha =1$ regime. The agreement improves as $k_Fa$ increases and the system becomes more dilute.}\label{fig:parity}
\end{figure}

It is interesting to compare the accuracy of the two-band model to the exact four-band solution. Several recent papers\cite{pientka:2013:1,pientka:2014:1,weststrom:2015:1} have considered a similar chain with a helical spin configuration in an intrinsically 3D system rather than the ferromagnetic planar version with Rashba SOC that is the focus of this article. As seen in Ref.~\onlinecite{weststrom:2015:1}, the two-band and four-band models are in fairly good agreement in the helical system; however, in the ferromagnetic case the differences of the models are more pronounced.  While the two-band and four-band models indeed agree in the deep-dilute limit $\alpha\approx1$, $k_Fa\gg1$, the convergence is much slower than in the helical 3D model. In order to compare the two models, we plot the $\mathbb{Z}_2$ phases of the four- and two-band models, using equations (\ref{eq:teemutrick}) and (\ref{eq:teemutrick2b}), in Fig.~\ref{fig:parity}. Even though the two-band model captures the behaviour reasonably in the dilute regime $k_Fa > 10\pi$, it is evident that non-negligible departures remain. In the dense regime $k_Fa\lesssim 4\pi$ the differences become qualitatively significant and the deep-impurity approximation starts to break down.

\begin{figure}
\includegraphics[width=0.9\linewidth]{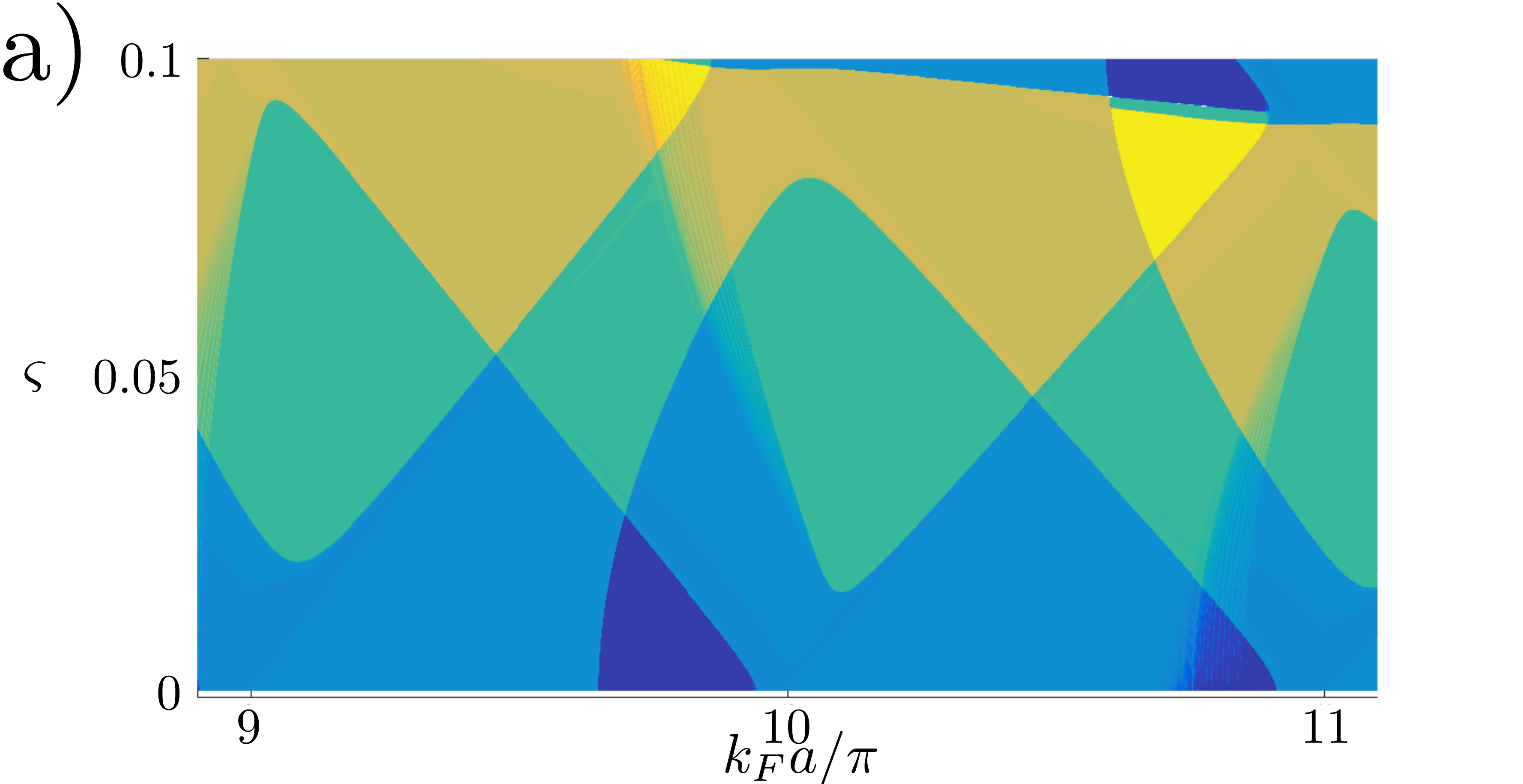}\\
\includegraphics[width=0.9\linewidth]{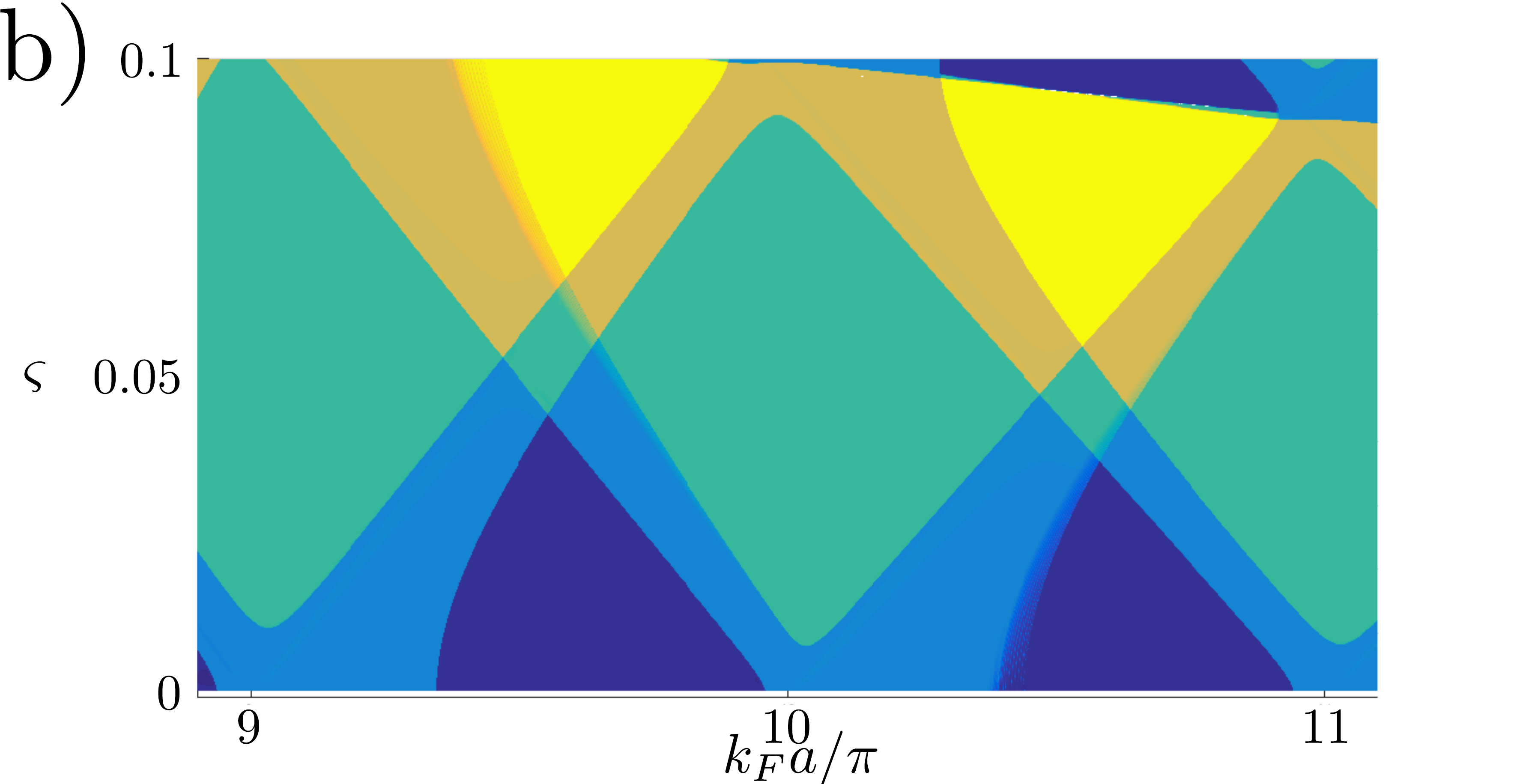}\\
\includegraphics[width=0.9\linewidth]{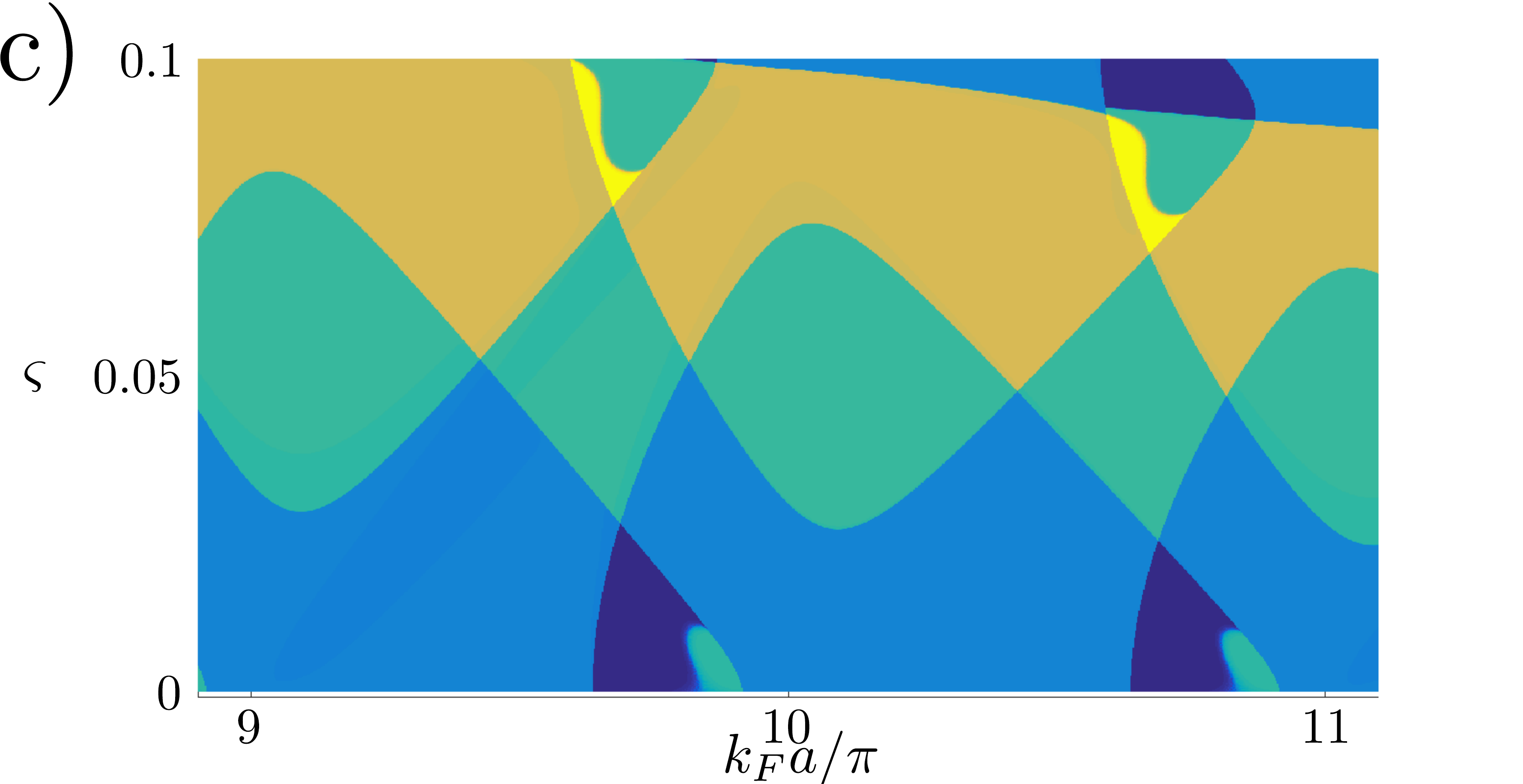}\\
\caption{ Topological phase diagram plotting the winding number as a function of $k_Fa$ and $\varsigma$. The color scheme is the same as Fig.~\ref{fig:winding}. a) Parameters used are $\alpha = 1$, $\xi_0 = 50a$ b) Same diagram, but $\alpha = 0.85$. The trivial regions at the center and edges of the figure are larger than in a), but so are the $|\mc N| = 2$ phases, even at the expense of some trivial regions. c) Topological phase diagram of the two-band model. Parameters used are $\alpha = 1$, $\xi_0 = 5a$; compare Fig.~1 in Ref.~\onlinecite{brydon:2015:1}, which plots the parity of the invariant for the same parameters.
}\label{fig:winding_varsigma}
\end{figure}

Having analysed the impact of the parameter $\alpha$, we move on to consider the behaviour of the topological phases as a function of the normalised SOC strength $\varsigma=m\alpha_R/k_F$. In Fig.~\ref{fig:winding_varsigma} we plot the winding number as a function of $k_Fa$ and $\varsigma$ for different values of $\alpha$. While the trivial regions do grow when the single-impurity bound state is moved away from the gap centre $\alpha=1$, for the parameters chosen here, the $\mc N = \pm 2$ phases grow as well; hence, lowering $\alpha$ in this case makes even parity more likely. While the linearised model may not quantitatively agree with the four-band model, its phase diagram shows many of the same characteristics. To make contact with previous work, we present in Fig.~\ref{fig:winding_varsigma}~c) a topological phase diagram similar to that of Fig.~1 in Ref.~\onlinecite{brydon:2015:1}.  The evaluation of the winding number reveals a richer structure of topological regions as well as entirely new non-trivial regions compared to the $\mathbb{Z}_2$ classification presented in that work.

\section{Majorana bound states } \label{MBS}
In the previous section we analysed the topological properties of the system, finding values for the winding number between $-2$ and $+2$. In practice, realisation of these phases may be limited by factors such as system size and temperature. In this section we will study the energy scales and spatial decay of the Majorana bound states in finite chains with open boundary conditions. This is carried out by numerically solving the $4N\times4N$ NLEVP (\ref{eq:nlevp}) in real space. 
\begin{figure*}
\begin{centering}
\includegraphics[width=0.4\linewidth]{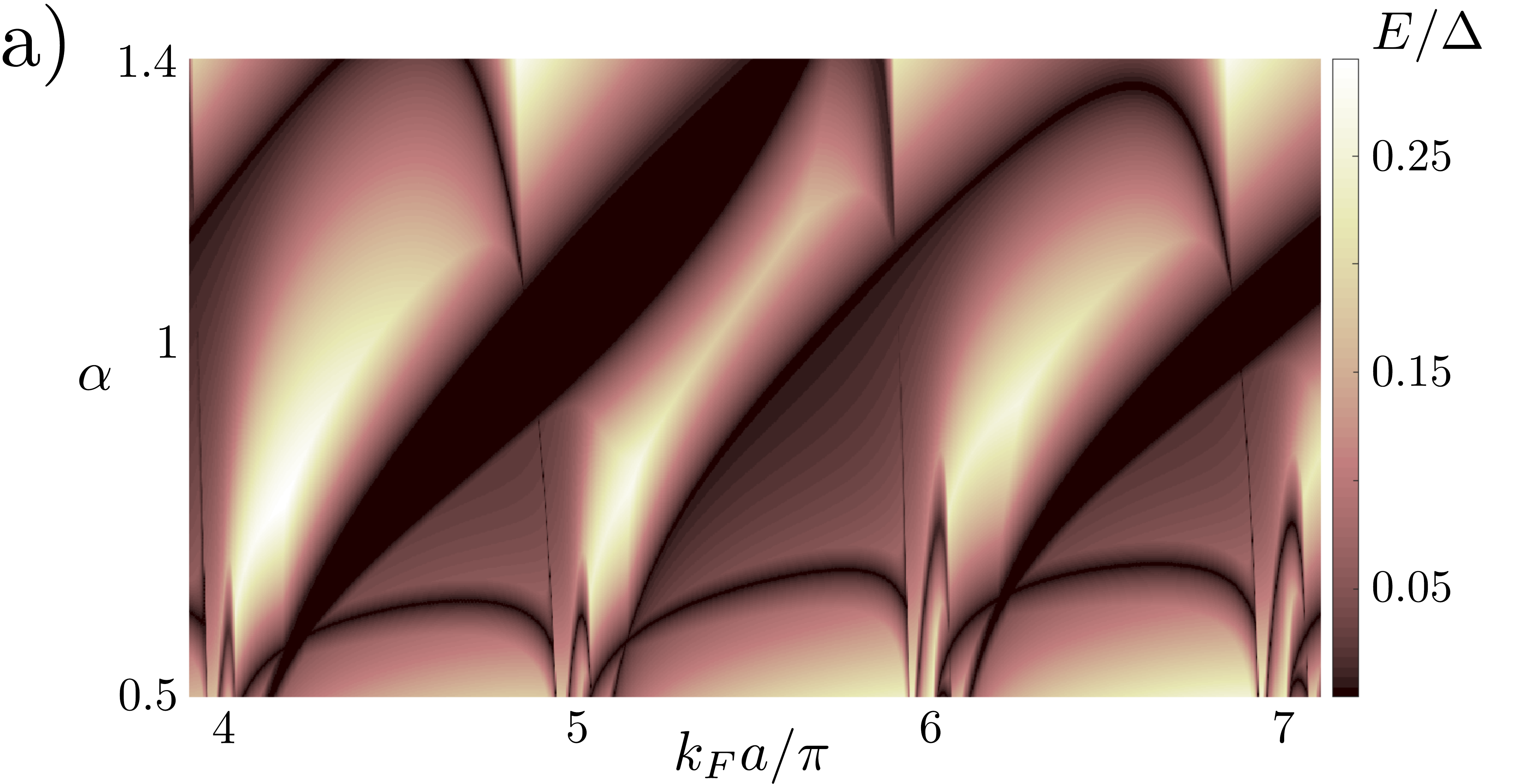}
\includegraphics[width=0.27\linewidth]{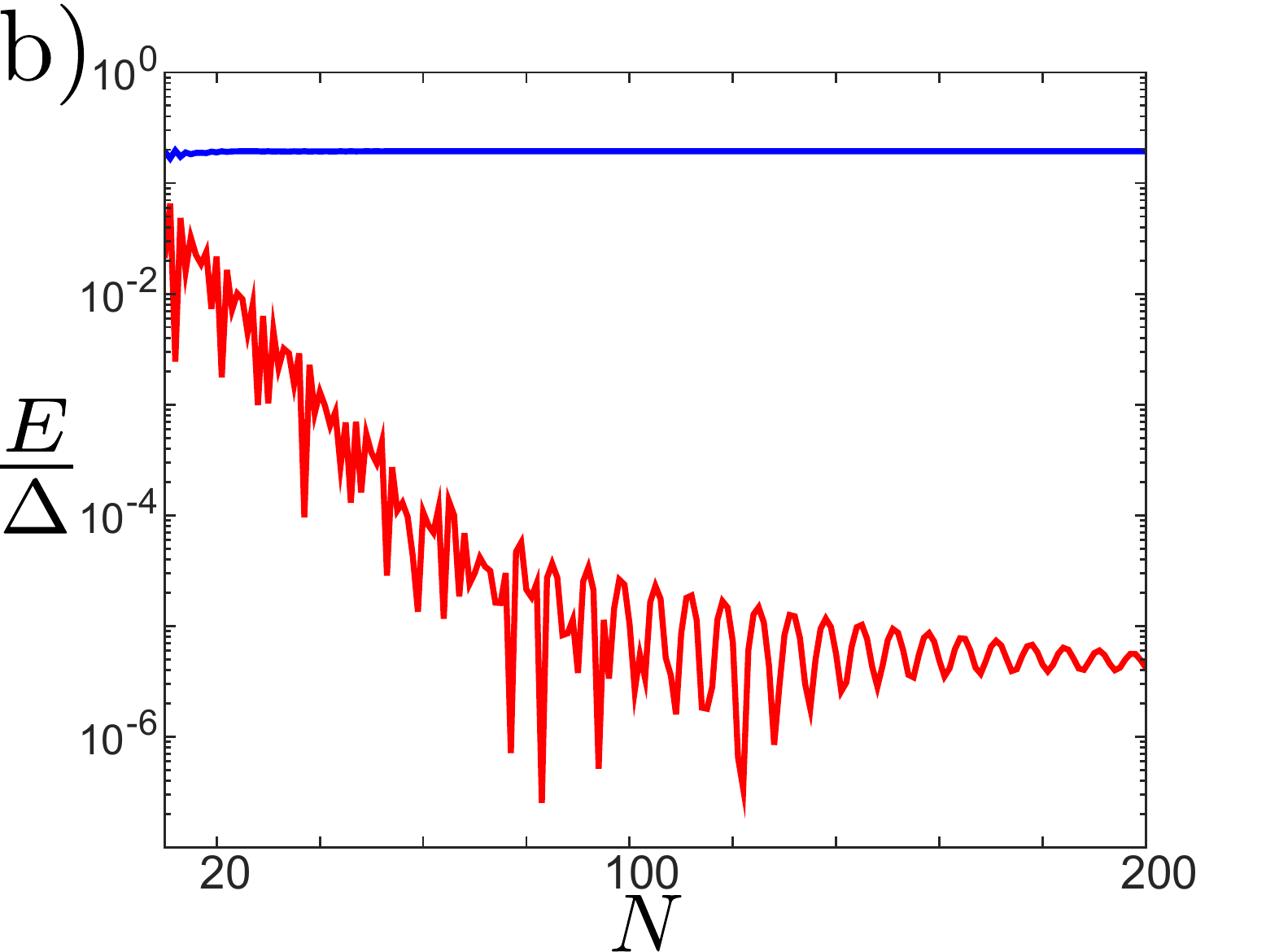}
\includegraphics[width=0.27\linewidth]{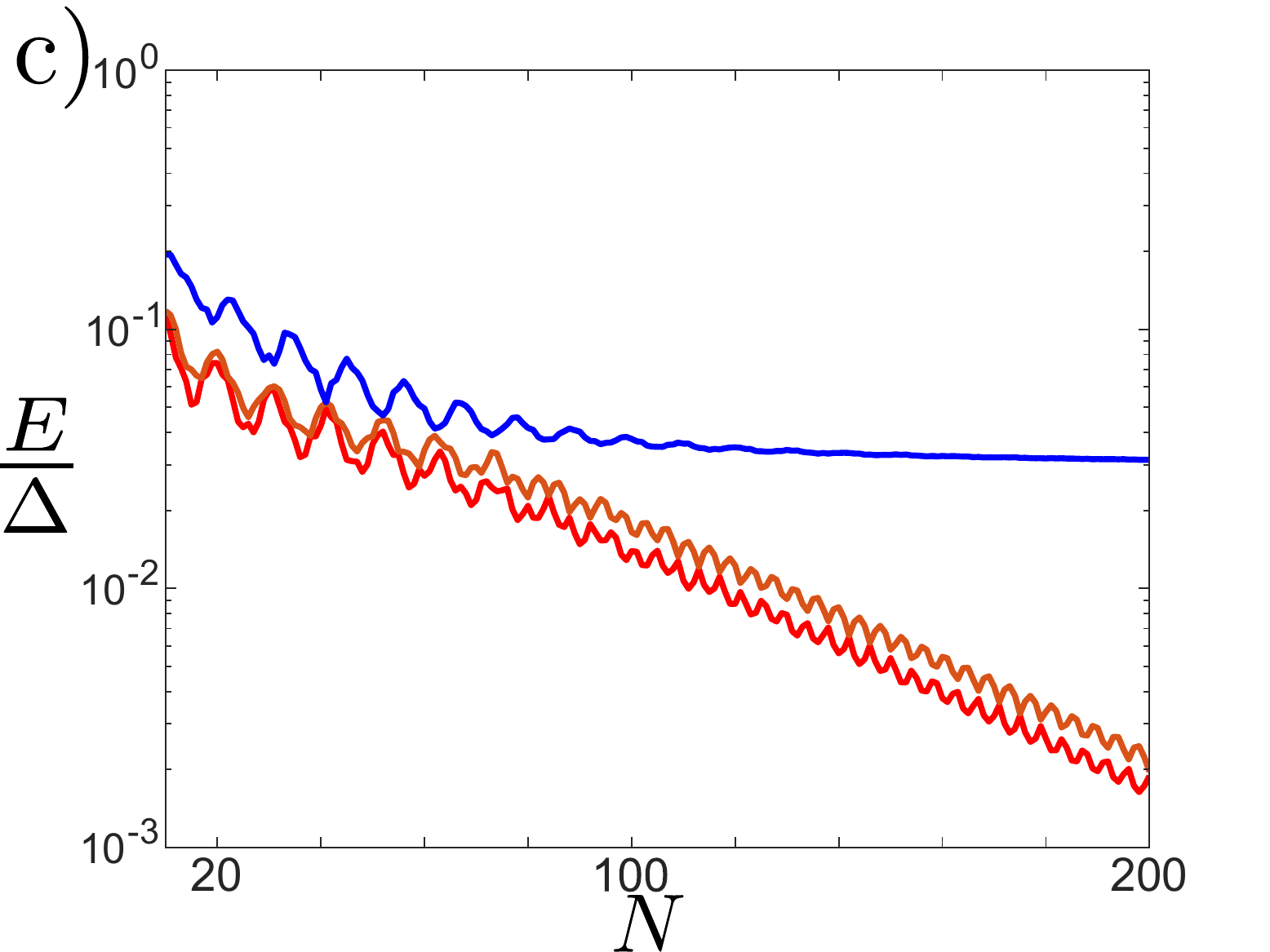}
\end{centering}
\caption{a) Bulk energy gap as a function of $k_Fa$ and $\alpha$. Parameters used are $\varsigma = 0.01$ and $\xi_0 = 50a$. The black lines correspond to topological phase transitions; the wide black zones are areas of low bulk gap. The corresponding winding number diagram is illustrated in Fig.~\ref{app:fig:xidiagram} in Appendix B. b) Dependence of bulk and MBS energy on the length of the chain $N$ in the $\mc N = 1$ phase. Parameters used are $k_Fa=20$, $\alpha=1$, $\varsigma = 0.01$, $\xi_0 = 50a$. c) Dependence of bulk and MBS energy on the length of the chain in the $\mc N = 2$ phase. Parameters used are $k_Fa = 5.7\pi$, $\alpha = 0.8$, $\varsigma = 0.01$, $\xi_0 = 50a$.}\label{fig:gaps}
\end{figure*}

A robust topological phase requires the existence of a well-defined energy gap between the MBSs at zero energy and the bulk states. In the limit of an infinite system, this gap can be straightforwardly calculated using Eq.~(\ref{eq:energy}). In Fig.~\ref{fig:gaps} a) we have plotted the energy gap of the system as a function of $k_Fa$ and $\alpha$. Excepting an area of low gap size around certain $k=0$ transitions, the energy gap is generally of the order of $0.1\Delta$, reaching values around $0.25\Delta$ at the center of the $|\mc N| = 1$ region; assuming a gap size similar to the $1.36$ meV  seen in the experiment on Pb surface in Ref.~\onlinecite{nadj-perge:2014:1}, this corresponds to temperatures $\approx 4$ K. While the parameters for a realistic system may not reach this maximum it is clear that there are wide regions where the gap size is large compared to temperatures that can realistically be reached in experimental setups. The $|\mc N| = 2$ phases generally have a much lower bulk gap, often by an order of magnitude. For finite systems the energy gap and the MBS splitting will depend on the chain length. In Fig.~\ref{fig:gaps} b) and c) we have plotted the gap and MBS energy as a function of chain length in the $|\mc N| =1$ and $|\mc N| = 2$ region, respectively. In the $|\mc N| = 1$ region the dependence of the energy gap on the chain length is negligible, and for relatively short chains with $N\lesssim50$ the MBS spitting is two orders of magnitude smaller than the energy gap.  The oscillating splitting of the MBSs localised at each end of the wire will first go down exponentially after which it settles on a slower algebraic suppression as discussed previously for wavefunctions in Ref.~\onlinecite{pientka:2014:1}. The algebraic tail of the MBS splitting could, in principle, be harmful for quantum information applications, though the absolute value of the splitting is orders of magnitude smaller than the gap. For the $|\mc N| = 2$ phase the energy gap is first reduced by an increasing chain length before converging to a constant value. Observation of the double MBSs at each end is limited by the large splitting seen in short chains, hence requiring larger system sizes and lower temperatures.

\begin{figure}
\includegraphics[width=0.48\linewidth]{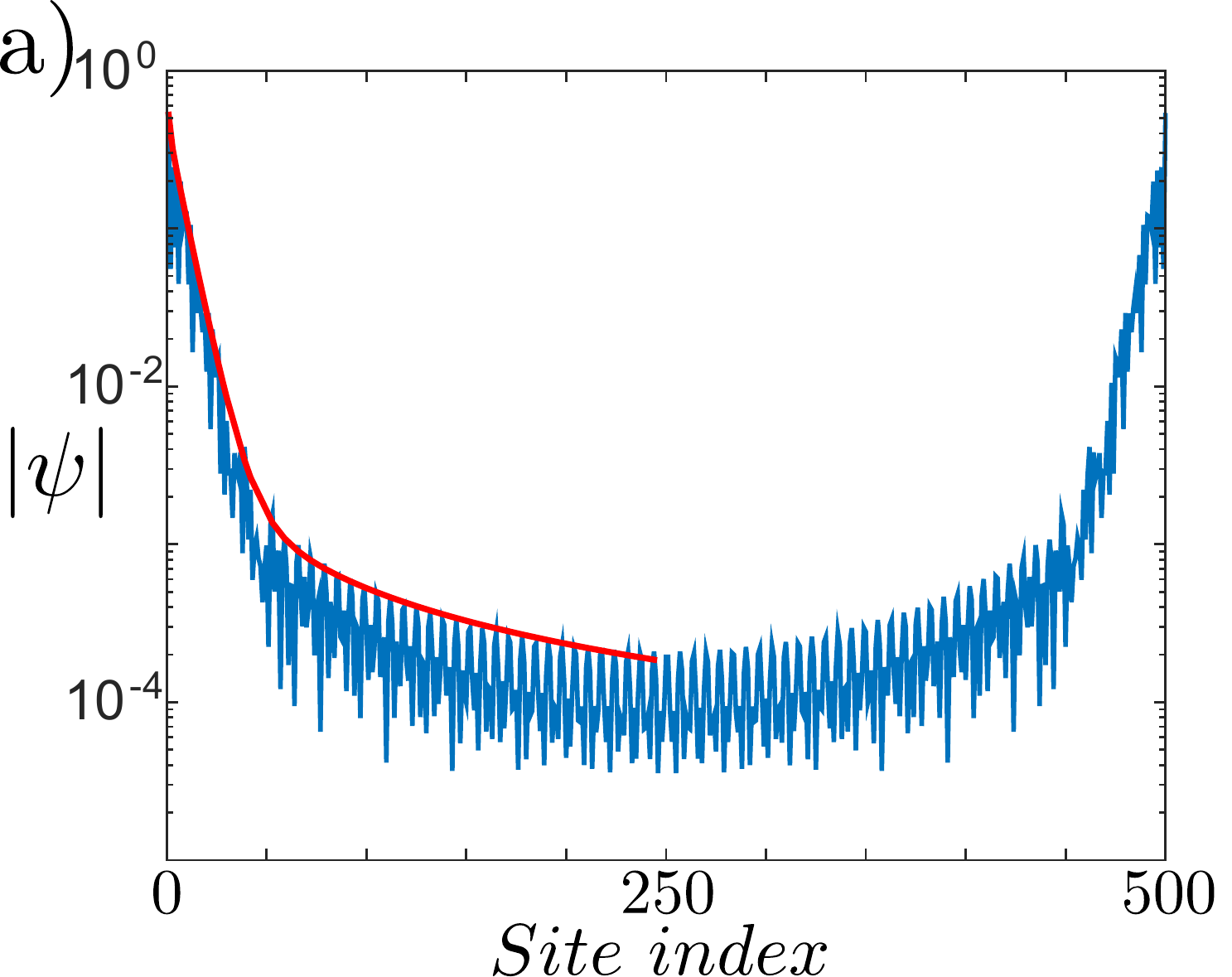}
\includegraphics[width=0.48\linewidth]{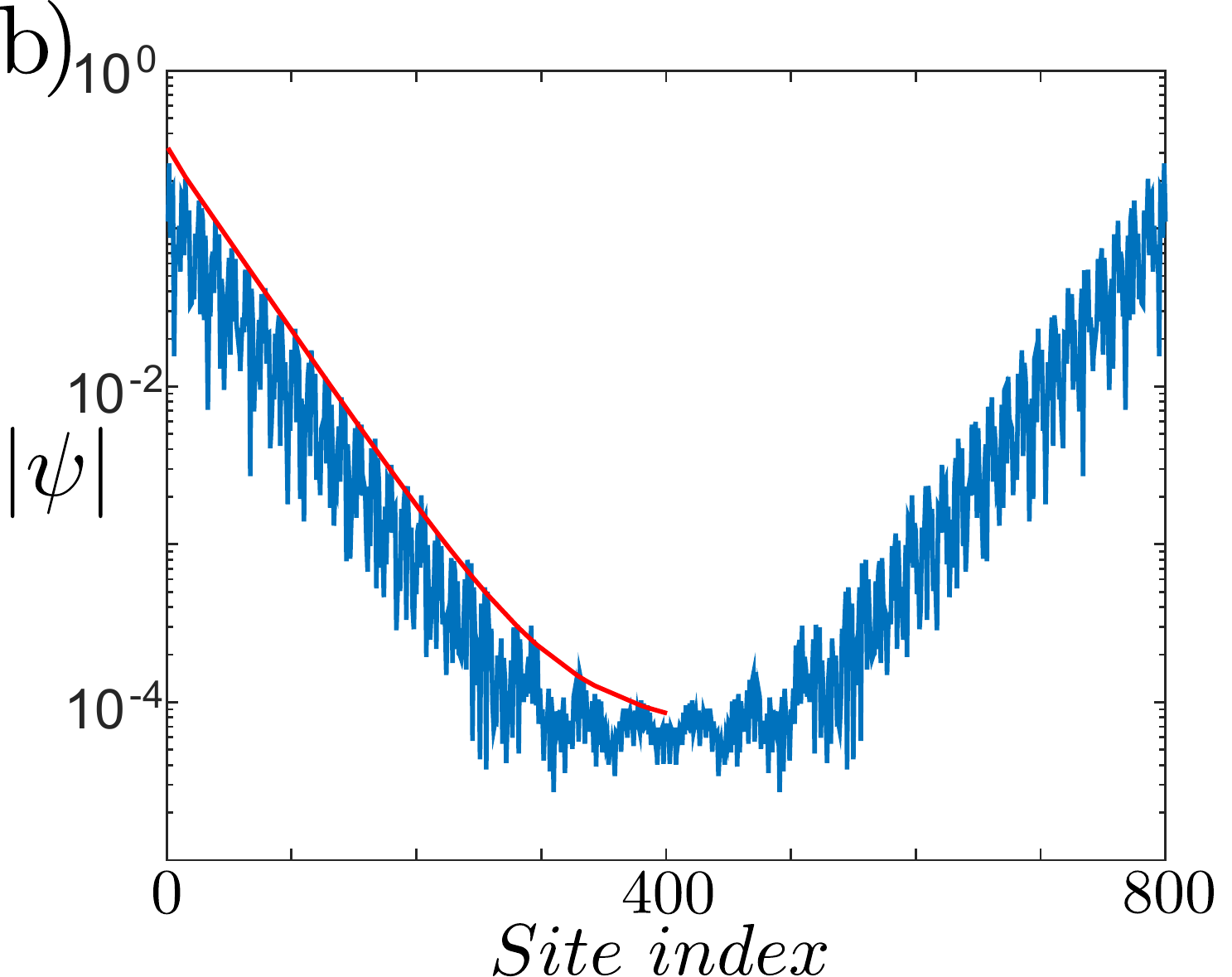}
\caption{Spatial decay of Majorana wavefunctions. a) Wavefunction in $\mc N = 1$ phase. Parameters used are $N = 500$, $k_Fa = 20$, $\alpha = 1$, $\varsigma = 0.01$, $\xi_0 = \infty$. The fitted function is $f(n) \approx 0.43e^{-0.16n} + 6.84n^{-1}\ln^{-2}(n/0.0011)$. b) Wavefunction in $\mc N = 2$ phase. Parameters used are $N = 800$, $k_Fa = 17.74$, $\alpha = 0.73$, $\varsigma = 0.01$, $\xi_0 = \infty$. The fitted function is $f(n) \approx 0.31e^{-0.026n} + 1.01n^{-0.79}\ln^{-2}(n/0.07)$.}\label{fig:decay}
\end{figure}

Besides the splitting behaviour we are also interested in the spatial decay of the MBS wavefunctions. This was previously analysed in the case of a helical chain by Pientka \textit{et al.}\cite{pientka:2014:1} An analytical solution of this problem is beyond the scope of this work, but numerical parameter fitting indicates that the envelope of the wavefunctions decays in a similar manner as those found in the helical chain. As seen in Fig.~\ref{fig:decay}, the decay is exponential over short distances but over longer distances (but still $<\xi_0$) takes a character of $x^{-\beta} \ln(x/x_0)^{-\gamma}$, where $\beta$, $\gamma \in \mathbb R_+$. This decay of the wavefunctions reflect the behaviour of the MBS energy splitting which has the same origin. In general, the MBSs seem to be more localised in the $|\mc N| = 1$ phases, while the $|\mc N| = 2$ phase features stronger finite-size effects which are also apparent in Fig.~\ref{fig:gaps}.

\section{Discussion and outlook} \label{summ}
Motivated by recent experimental developments in topological chains, we studied the case of a ferromagnetic chain of adatoms embedded on a superconductor with Rashba spin-orbit coupling. Building upon previous works limited to the deep magnetic impurity regime, we formulated a theory that can address 1D Shiba bands at arbitrary subgap energies. This allowed us to make a number of novel observations regarding  the topological states in the chain.

We found that the system, in accordance with its BDI classification which allows a $\mathbb{Z}$-valued topological invariant, supports five different topological phases, some of which have hitherto not been reported in this setup. These phases may support zero, one or two Majorana end states and their interfaces may support up to four Majorana bound states. Our analysis revealed that energy gaps of the most robust phases may be as high as $0.2\Delta$ and that the decoupled impurity energy does not need to be fine-tuned close to the gap centre for obtaining robust topological phases. In the dense-chain limit\cite{peng:2015:1,li:2014:2}, where the orbitals of the magnetic atoms have a direct overlap, the mechanisms of topological phases are qualitatively quite different, bearing a strong resemblance to the proximity coupled nanowire physics. However, it is remarkable that robust topological gaps can be obtained also in the studied Shiba limit with bound state energies well away from the gap centre. The sources and effects of disorder in the long-range hopping model are expected to be quite different from the conventional nanowire setting and are left for future studies. Considering the richness of the topological properties, and the fact that the system allows for local probing in STM, the studied system is one of the most promising platforms for studying topological superconductivity.    

The main obstacle for applications of topological superconductivity\cite{alicea:2011:1,li:2014:1} may be the slow spatial decay of the edge modes, which results in an algebraic energy splitting of the Majorana end states. However, already in relatively short wires the splitting could be orders of magnitude smaller than the bulk energy gap, as found in our numerical calculations. Our results are generally encouraging for the prospect of an experimental observation of topological superconductivity in the studied setup. 
\acknowledgments

The authors acknowledge the Academy of Finland for support.

\appendix
\numberwithin{equation}{section}

\section{Derivation of the NLEVP}
In this appendix we will derive the nonlinear eigenvalue problem for the four-band Rashba model. Our starting point is Eq.~(\ref{eq:postBdG})
\begin{equation}
\begin{split}
\left[E-\left(\xi_{\vec k} + \alpha_R(k_y\sigma_x - k_x\sigma_y)\right)\tau_z - \Delta\tau_x\right]\Psi(\vec r) \\= - J\sum_j S\sigma_z\delta(\vec{r} - \vec{r}_j)\Psi(\vec r_j)
\end{split},
\end{equation}
which we then Fourier transform, giving us
\begin{equation}
\begin{split}
\left[E-\left(\xi_{\vec k} + \alpha_R(k_y\sigma_x - k_x\sigma_y)\right)\tau_z - \Delta\tau_x\right]\Psi_{\vec k} \\= - J\sum_j e^{-i \vec k \cdot \vec r_j}S\sigma_z\Psi(\vec r_j).
\end{split}
\end{equation}
Multiplying the equation by the inverse of the matrix on the left-hand side, and transforming back to real space yields
\begin{equation}\label{app:rashba3}
\left(1 + J_E(0)S\sigma_z\right)\Psi(x_i) = -\sum_{j\neq i} J_E(x_i - x_j)S\sigma_z\Psi(x_j).
\end{equation} with the identification
\begin{widetext}
\begin{equation}
J_E(\vec r) = JS\int \frac{d\vec k}{(2\pi)^2} e^{i\vec k\cdot \vec r} [E - (\xi_{\vec k} + \alpha_R (k_y\sigma_x - k_x\sigma_y))\tau_z  - \Delta\tau_x]^{-1}.
\end{equation}
\end{widetext}
By defining
\begin{equation}
\vec M_\pm = \frac{E + \xi_\pm\tau_z + \Delta\tau_x}{E^2 - \xi_\pm^2 - \Delta}\left(1\pm \frac{k_y}{k}\sigma_x \mp \frac{k_x}{k}\sigma_y\right),
\end{equation}
where $k = \sqrt{k_x^2+k_y^2}$ and $\xi_\pm = \xi_{\vec k} \pm \alpha_R k$, we split $J_E$ into two helicities corresponding to each sign,
\begin{equation}
J_E(\vec r) = JS\int \frac{d\vec k}{(2\pi)^2} e^{i\vec k\cdot \vec r} (\vec M_+ + \vec M_-).
\end{equation}
We assume that the wavevector $k$ is centered around the point $\xi_\pm \approx 0$, or, equivalently, in the neighbourhood of the Fermi surface for both helicities. In spherical coordinates, we can then substitute $k$ for $\xi_\pm$ in the integrals as appropriate. This gives us four different integrals to calculate, namely
\begin{equation}
\begin{split}
I^\pm_1(r) &= \frac{N_\pm}{(2\pi)^2}\int_0^{2\pi}d\theta\int_{-\infty}^{\infty}d\xi\frac{\xi e^{ik_\pm(\xi)r\cos\theta}}{E^2-\xi^2 -\Delta^2}\\
I^\pm_2(r) &= \frac{N_\pm}{(2\pi)^2}\int_0^{2\pi}d\theta\int_{-\infty}^{\infty}d\xi\frac{\xi e^{i\theta + ik_\pm(\xi)r\cos\theta}}{E^2-\xi^2 -\Delta^2}\\
I^\pm_3(r) &= \frac{N_\pm}{(2\pi)^2}\int_0^{2\pi}d\theta\int_{-\infty}^{\infty}d\xi\frac{ e^{ik_\pm(\xi)r\cos\theta}}{E^2-\xi^2 -\Delta^2}\\
I^\pm_4(r) &= \frac{N_\pm}{(2\pi)^2}\int_0^{2\pi}d\theta\int_{-\infty}^{\infty}d\xi\frac{ e^{i\theta + ik_\pm(\xi)r\cos\theta}}{E^2-\xi^2 -\Delta^2}.
\end{split}
\end{equation}
Here we have introduced $N_\pm = 1 \mp \varsigma/\sqrt{1+\varsigma^2}$ and $k_\pm(\xi) = k_F(\sqrt{1+\varsigma^2}\mp\varsigma) + \xi/v_F\sqrt{1+\varsigma^2}$, with the normalised spin-orbit coupling $\varsigma = m\alpha_R/k_F$. To make further progress, we restrict our attention to subgap energies, i.e. $E < \Delta$. For $r = 0$, all integrals but the third vanish trivially, and $I_3^\pm$ is simplified enough to be calculated using standard contour integral methods. For $r>0$ we must employ Bessel and Struve functions. Setting the chain to be along the $x$-direction, we can now rewrite Eq.~(\ref{app:rashba3}) into the form of Eq.~(\ref{eq:rashbaeq}) with
\begin{equation}
\begin{split}
J_E(x) = \tfrac{\alpha}{2}{\left[(I^-_1(x) + I^+_1(x))\tau_z +(I^-_2(x) - I^+_2(x))\tau_z\sigma_y \right]}\\
+\tfrac{\alpha}{2}{(E + \Delta\tau_x)}{\left[(I^-_3(x) + I^+_3(x)) + (I^-_4(x) - I^+_4(x))\sigma_y\right]},
\end{split}
\end{equation}
where
\begin{widetext}
\begin{equation}
\begin{split}
I^\pm_1(r) &= N_\pm\IM\left[J_0((k_{F,\pm} + i\xi_E^{-1})|x|))+ i H_0((k_{F,\pm} + i\xi_E^{-1})|x|))\right]\\
I^\pm_2(r) &= -iN_\pm\sgn(x)\RE\left[iJ_1((k_{F,\pm} + i\xi_E^{-1})|x|))+ H_{-1}((k_{F,\pm} + i\xi_E^{-1})|x|))\right]\\
I^\pm_3(r) &= -\frac{N_\pm}{\sqrt{\Delta^2-E^2}}\RE\left[J_0((k_{F,\pm} + i\xi_E^{-1})|x|))+ i H_0((k_{F,\pm} + i\xi_E^{-1})|x|))\right]\\
I^\pm_4(r) &= -i\frac{N_\pm}{\sqrt{\Delta^2-E^2}}\IM\left[iJ_1((k_{F,\pm} + i\xi_E^{-1})|x|))+ H_{-1}((k_{F,\pm} + i\xi_E^{-1})|x|))\right].
\end{split}
\end{equation}
\end{widetext}
Here we have used the previously mentioned Bessel ($J_{0, 1}$) and Struve ($H_{0, -1}$) functions along with the two Fermi momenta $k_{F,\pm} = k_F(\sqrt{1+\varsigma^2}\mp\varsigma)$ and the quantity $\xi_E = \xi_0/\sqrt{1 - E^2/\Delta^2}$. The problem can be further simplified by working in the basis of the eigenstates of $\tau_x\sigma_z$, which correspond to the eigenstates of the single-impurity problem. With the shorthand $\ket{\tau_x\sigma_z}$, where for example $\ket{+\uparrow} = \ket{+}_{\tau_x}\otimes\ket{\uparrow}_{\sigma_z}$, the spinor in the new basis is
\begin{equation}\label{app:eq:basis}
\Psi_j = (\brakets{{+\uparrow}}{\Psi_j}\ \brakets{{-\downarrow}}{\Psi_j}\ \brakets{{+\downarrow}}{\Psi_j}\ \brakets{{-\uparrow}}{\Psi_j})^T.
\end{equation}
In the transformed basis, we obtain a $4N\times 4N$ nonlinear eigenvalue problem for a chain of length $N$. By defining the matrices
\begin{equation}
\begin{split}
a_{ij} &= -\frac{\sqrt{\Delta^2-E^2}}{2m}(I_3^-(x_{ij}) + I_3^+(x_{ij})) + \delta_{ij}\\
b_{ij} &= - \frac{i}{2m}(I_2^-(x_{ij}) - I_2^+(x_{ij}))\\
c_{ij} &= -i\frac{\sqrt{\Delta^2-E^2}}{2m}(I_4^-(x_{ij}) - I_4^+(x_{ij}))\\
d_{ij} &= - \frac{1}{2m}(I_1^-(x_{ij}) + I_1^+(x_{ij}))
\end{split}
\end{equation}
we can transform Eq.~(\ref{eq:rashbaeq}) into the NLEVP in Eq.~(\ref{eq:nlevp}).

\section{Validity of the energy-independent coherence length approximation}

The purpose of this appendix is to give some motivation as to why using $\xi_0$ instead of $\xi_E$ in the four-band model is an excellent approximation. We first note that for a given $\xi_0$, we have $\xi_E \geq \xi_0$, so it is evident that the real energy must lie within the interval $[f_{\inf}, f_{\sup}]$, where $f_{\inf/\sup}$ stands for the infimum/supremum in the interval $\xi \in [\xi_0, \infty)$ for the function $f$ defined by $E = \Delta f(\xi_E)$, i.e. the right-hand side of the expression for the energy in Eq.~(\ref{eq:energy}). This, however, does not narrow the error down too much as $f(\xi)$ can vary greatly between $\xi = \xi_0$ and $\xi = \infty$.

One way to approach the problem is through the use of contractions. A contraction $g: \mathbb{R}\to\mathbb{R}$ is a Lipschitz continuous function, whose Lipschitz constant $K_g < 1$. In other words, if $g$ is a contraction and $\mathbb{R}$ is equipped with the usual metric, we have
\begin{equation}\label{app:eq:lipschitz}
|g(x) - g(y)| \leq K_g|x-y|,
\end{equation}
where $x,y\in\mathbb{R}$ and $0 < K_g < 1$. If we now define the iteration of a function for some $x\in\mathbb{R}$ as $x_0 = x, x_1 = g(x),\cdots,x_n = g(x_{n-1})$, we immediately get that 
\begin{equation}
|x_n - x_{n+1}| \leq K^n_g |x_0 - x_1|.
\end{equation}
Furthermore, this tells us that for any positive integers $p$ and $n$
\begin{equation}
\begin{split}
|x_n - x_{n+p}| \leq \sum_{m=0}^{p-1}|x_{n+m} - x_{n+m+1}| \\ \leq \sum_{m=0}^{p-1}K_g^{n+m}|x_0-x_1| = \frac{K_g^n(1-K^p_g)}{1-K_g}|x_0-x_1|,
\end{split}
\end{equation}
where we in the first step applied the triangle inequality. Since this is valid for any $p$, we can take the limit $p\to\infty$, resulting in
\begin{equation}
|x_n - x_*| \leq \frac{K_g^n}{1-K_g}|x_0-x_1|.
\end{equation}
Here $x_*$ is the so called fixed point defined by $x_* = g(x_*)$. For contractions, this fixed point is known to exist and be unique according to the Banach fixed point theorem.
\begin{figure}
\includegraphics[width=0.48\linewidth]{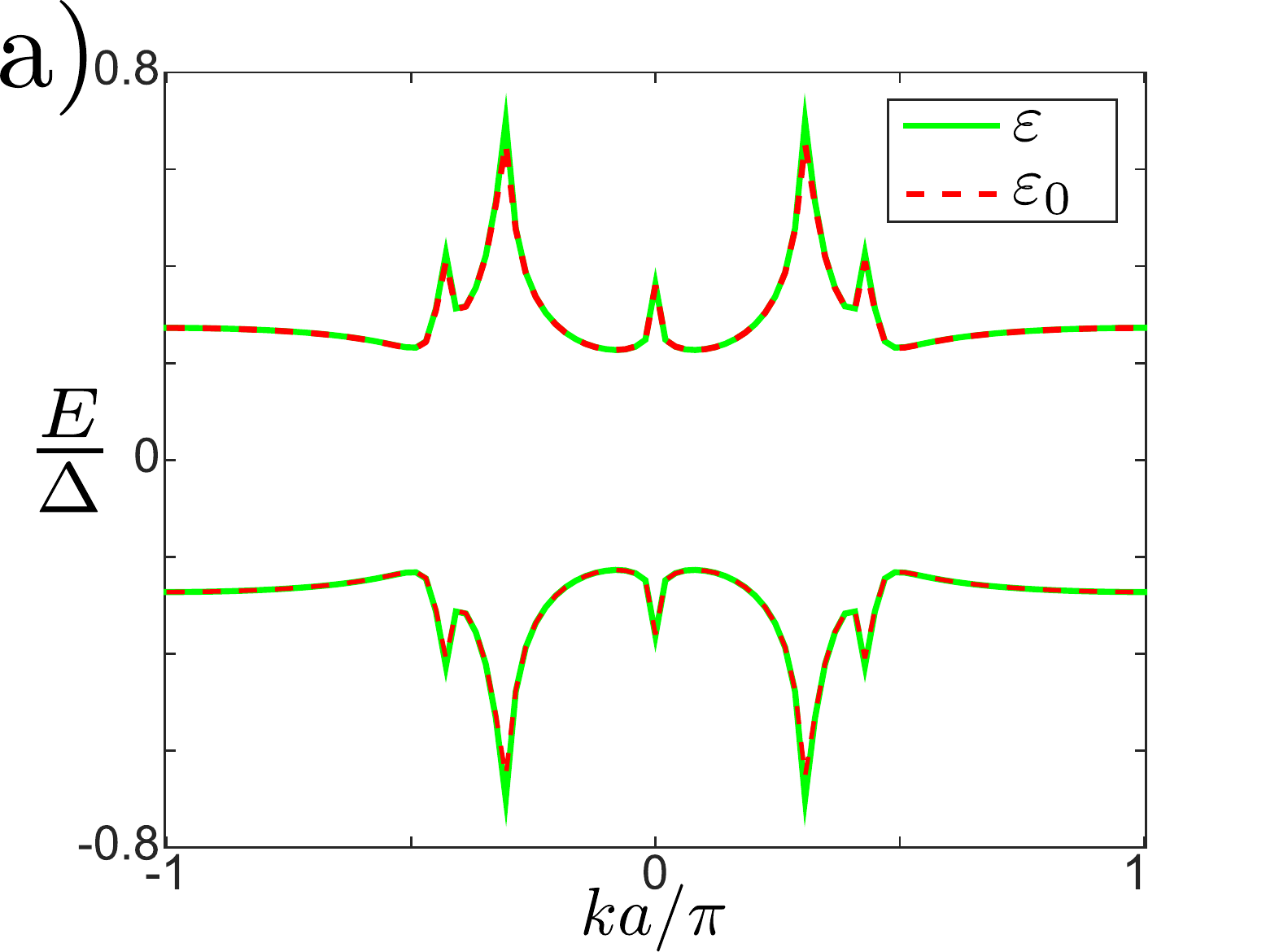}
\includegraphics[width=0.48\linewidth]{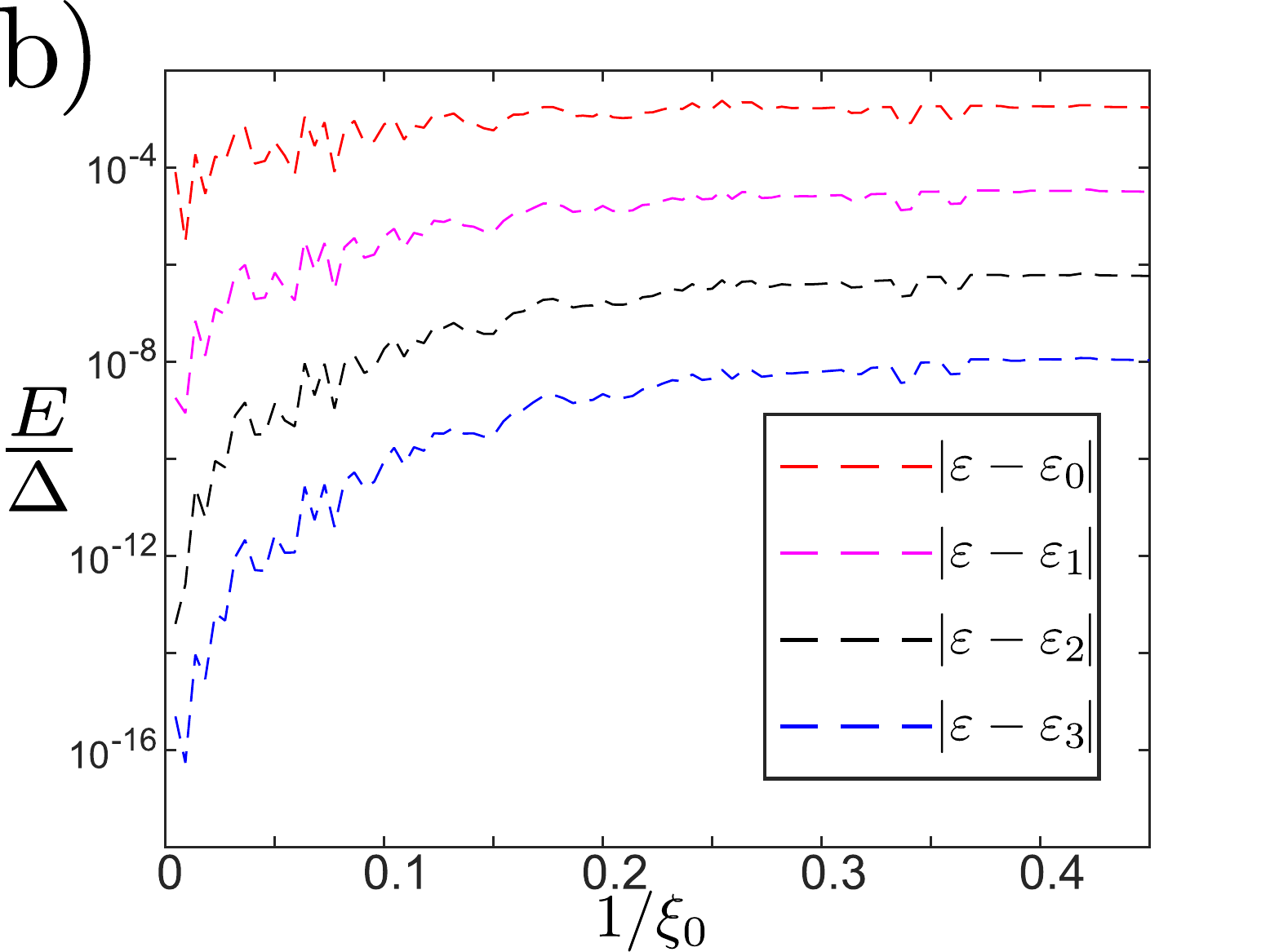}
\caption{a) Energy as a function of $k$ for the first approximation $\epsilon_0$ as well as the numerically exact fixed-point solution $\epsilon$ of Eq.~(\ref{eq:energy}). Parameter values are $k_Fa = 20$, $\alpha = 1$, $\varsigma = 0.01$, $\xi_0 = 50a$. Iterations are omitted here as they would be indistinguishable from the fixed-point solution.  b) Difference in energy between the solution of the transcendental equation and several iterations of the solution for $\epsilon_n$, plotted as a function of the inverse coherence length at zero energy.  Parameters used are $k_Fa = 20$, $\alpha=1$, $\varsigma=0.01$. We have arbitrarily chosen the point $k=\pi/5$ for the graphical presentation.}\label{app:fig:forcedxi}
\end{figure}

If we now introduce the dimensionless parameters $\epsilon \equiv E/\Delta$ and $\chi \equiv a/\xi_0$, we can rewrite the equation for the energy of a fixed point in reciprocal space as
\begin{equation}
\epsilon = f(\chi\sqrt{1-\epsilon^2}) \equiv h(\epsilon).
\end{equation}
Evidently, the energy satisfying this equation is the fixed point of $h$. If $h$ is then also a contraction, we can apply the machinery developed in the previous paragraph. Through the mean-value theorem, we know that for any $x < z\in[0,1)$ there exists a $y\in[x,z]$ such that
\begin{equation}
|h(x) - h(z)| = |x-z|\chi\frac{|y|}{\sqrt{1-y^2}}\frac{df(x)}{dx}\bigg\vert_{x = \chi\sqrt{1-y^2}}.
\end{equation}
\begin{figure}
\includegraphics[width=0.98\linewidth]{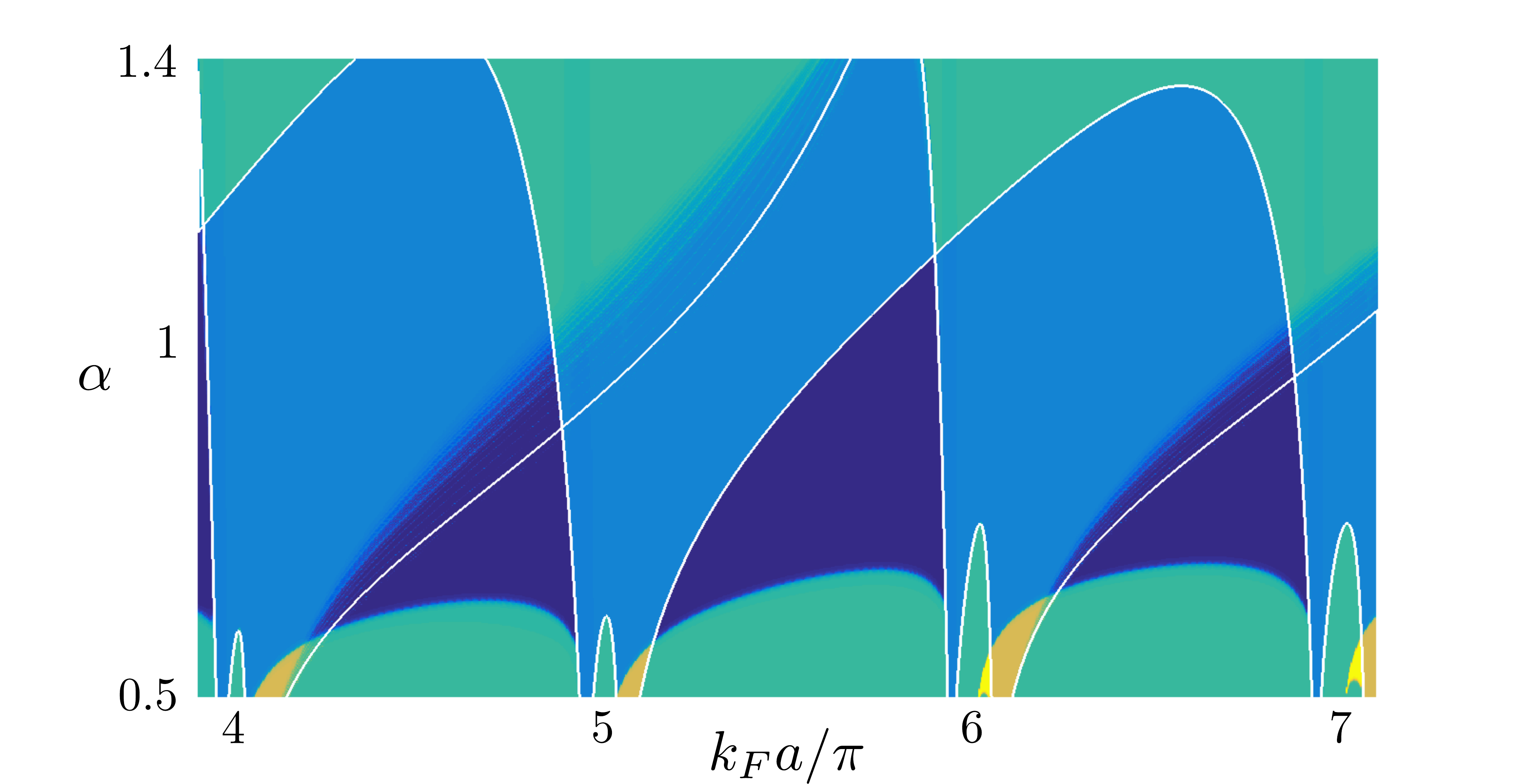}
\caption{Topological phase diagram obtained by calculating the winding number using the $\xi\approx \xi_0$ approach. The white line plots the border between regions where the topological invariant is of different parity, as calculated using Eq.~(\ref{eq:teemutrick}); this equation is valid for arbitrary coherence lengths. Parameters used are $\alpha_R = 0.01$, $\xi_0 = 50a$.}\label{app:fig:xidiagram}
\end{figure}
Comparing to Eq.~(\ref{app:eq:lipschitz}), then, if
\begin{equation}
K_h = \chi\max_{y\in [f_{inf}, f_{sup}]}\left(\frac{|y|}{\sqrt{1-y^2}}\frac{df(x)}{dx}\bigg\vert_{x = \chi\sqrt{1-y^2}}\right) < 1,
\end{equation}
$h$ becomes a contraction. Usually $K_h$ is of the order of $10^{-2}-10^{-1}$ regardless of the chosen $k$ point, but due to the complicated form of $f$ this has to be checked for each case separately. Nevertheless, we can finally conclude that
\begin{equation}
|\epsilon_0-\epsilon| \leq \frac{K_h}{1-K_h}|\epsilon_0-\epsilon_1|,
\end{equation}
where $\epsilon_0 = h(0),\ \epsilon_1 = h(\epsilon_0),$ and $\epsilon = h(\epsilon)$ is the actual normalised energy. Note that the approximation $\xi_E \to \xi_0$ corresponds to $\epsilon\to\epsilon_0$. An example of the effect of this approximation is seen in Fig.~\ref{app:fig:forcedxi} a), where the first approximation $\epsilon_0(k)$ and the fixed-point solution $\epsilon(k)$ are plotted in the same figure. The iterations converge towards the exact solution very rapidly, but, as seen in the figure, the first approximation using $\xi_0$ is already reasonably close. A comparison of the differences between iterations for a selected $k$ point is seen in subfigure b); the initial error is minor, and each iteration reduces it by several orders of magnitude. As expected the errors are lowest at high coherence lengths; however, even at low coherence lengths, error for the first iteration remains lower than $10^{-2}\Delta$.

Another way to test the accuracy of our approximation, is to compare the topological phase diagram calculated using Eq.~(\ref{eq:finalwinding}) to the phase diagram of the $\mathbb{Z}_2$ invariant found using  Eq.~(\ref{eq:teemutrick}). Note that the former method relies on the approximation, while the latter is inherently equally valid for any coherence length.  As even the quantitative differences between iterations are generally negligible, the qualitative effects on the topological properties of the system are minor. In Fig.~\ref{app:fig:xidiagram} we see these two methods in superposition. The regions in which the two methods disagree coincide with areas where the gap size is extremely low (see Fig.~\ref{fig:gaps}~a)). The apparent disagreement between the two phase diagrams is hence due to unreliability in the numerical calculation of the winding number, rather than the $\xi_0$ approximation. Physically this is of little relevance, as the gap size in that are is too low for observation of topological effects of the type studied here.

\section{The Winding Number formula}
The topological invariant for the one-dimensional FM Shiba chain can be calculated using Eq.~(\ref{eq:winding}). This is straightforward for the linearised two-band model, but the four-band NLEVP is not formulated in terms of a Hamiltonian operator. As mentioned in the main text, this can be circumvented by defining a topologically equivalent
 flat-band version of the effective Hamiltonian, $\tilde H = \hat P_+ - \hat P_-$. Inserting this expression in Eq.~(\ref{eq:winding})  gives
\begin{equation}
\mathcal N = \frac{1}{4\pi i}\int_{-\pi/a}^{\pi/a} \mathrm{d}k\ \tr\left[\mathcal C (\hat{P}_+-\hat{P}_-)\partial_k(\hat{P}_+-\hat{P}_-)\right].
\end{equation}
We can progress by considering the chiral symmetry operator $\mc C$. The effect of $\mc C$ on the energy eigenstates is known to be $\mc C\ket{E_\pm} = \ket{E_\mp}$. It is straightforward to see that this also satisfies $\mc C \tilde H = - \tilde H \mc C$. Together with the properties of the trace, we obtain
\begin{equation}
\mathcal N = \frac{1}{2\pi i}\int_{-\pi/a}^{\pi/a} \mathrm{d}k\ \tr\left[\mathcal C \hat{P}_+\partial_k \hat{P}_+ -\hat{P}_+\mathcal C\partial_k  \hat{P}_+ \right].
\end{equation}
Integration by parts finally yields Eq.~(\ref{eq:finalwinding}) from the main text. 


\end{document}